\documentclass[prd,aps,twocolumn,floats,floatfix,superscriptaddress,preprintnumbers,showpacs]{revtex4}
\usepackage{amsfonts,amsmath,epsfig,amssymb,latexsym,eucal,color,graphicx,array,subfigure,bm}
\definecolor{red  }{rgb}{1,0,0}
\definecolor{blue }{rgb}{0,0,1}
\definecolor{green}{rgb}{0,1,0}
\begin{document}

\preprint{WU-AP/299/09}

%\vspace{1cm}

%---------------------------------------------------------------------%
\title{Measurement of the Kerr Spin Parameter by Observation\\of a Compact Object's Shadow}
%\vspace{1cm}
%---------------------------------------------------------------------%

\author{Kenta {\sc Hioki}}
\email{hioki@gravity.phys.waseda.ac.jp}
\address{Department of Physics, Waseda University, Okubo 3-4-1, Shinjuku, Tokyo 169-8555, Japan}
\address{Waseda Research Institute for Science and Engineering,
Okubo 3-4-1, Shinjuku, Tokyo 169-8555, Japan}
\author{Kei-ichi {\sc Maeda}}
\email{maeda@waseda.jp}
\address{Department of Physics, Waseda University, Okubo 3-4-1, Shinjuku, Tokyo 169-8555, Japan}
\address{Waseda Research Institute for Science and Engineering,
Okubo 3-4-1, Shinjuku, Tokyo 169-8555, Japan}

%---------------------------------------------------------------------%
%---------------------------------------------------------------------%
%---------------------------------------------------------------------%

\begin{abstract}
A black hole casts a shadow as an optical appearance because of its strong gravitational field.
We study how to determine the spin parameter and the inclination angle
by observing the apparent shape of the shadow, which is distorted mainly by those two parameters.
Defining some observables characterizing the apparent shape (its radius and distortion parameter),
we find that the spin parameter and inclination angle
of a Kerr black hole can be determined by the observation.
This technique is also extended to the case of a Kerr naked singularity.
\end{abstract}
\maketitle

%%%%%%%%%%%%%%%%%%%%%%%%%%%%%%%%%%%%%%%%%%%%%%%%%%%%%%%%%%%%%%%%%%%%%%%
%%%%%%%%%%%%%%%%%%%%%%%%%%%%%%%%%%%%%%%%%%%%%%%%%%%%%%%%%%%%%%%%%%%%%%%
%%%%%%%%%%%%%%%%%%%%%%%%%%%%%%%%%%%%%%%%%%%%%%%%%%%%%%%%%%%%%%%%%%%%%%%
\section{Introduction}
\label{sec:1}
%%%%%%%%%%%%%%%%%%%%%%%%%%%%%%%%%%%%%%%%%%%%%%%%%%%%%%%%%%%%%%%%%%%%%%%
%%%%%%%%%%%%%%%%%%%%%%%%%%%%%%%%%%%%%%%%%%%%%%%%%%%%%%%%%%%%%%%%%%%%%%%
It is widely believed that there exist black holes in the centers of 
many galaxies.
Sgr $\mathrm{A^\ast}$, which is the compact radio source at the center of the 
Milky Way, is highly likely to be a supermassive black hole.
In fact, the Newtonian orbital motion of the surrounding stars indicates the 
mass of the dark compact object 
at the Galactic center to be $M\sim 3.6\times 10^6 
M_\odot $~\cite{Eisenhauer:2005cv}.
Since the galaxies are rotating, it is very likely
that a black hole at the center of the galaxy also possesses a spin.
The analysis of the iron $K_\alpha$ emission line in the X-ray region 
indicates that the fine structure of the line spectra shows
the signature of the black hole spin~\cite{Tanaka:1995en,Reynolds:2002np}.
However, this is difficult to conclude because there is still some ambiguity and the result may
depend on the model of a gas inflow into a black hole.

New methodological development for measuring the spin of a black hole is
the grand challenge in the next generation of astronomy.
The direct observation of black holes by
future interferometers will become possible in the near future~\cite{Cash:2000exp,Hirabayashi:2005kc,Doeleman:2008qh,Doeleman:2008xq}.
In such an observation, a black hole may cast a shadow
in the sky as an optical appearance due to its strong gravity,
which we may call the shadow of the black hole. 
One has to exercise due care in the handling of the direct imaging of the shadow,
where a strong gravitational lensing effect plays a crucial role.
Though the theory of gravitational lensing has been well developed in the weak 
field approximation and has succeeded to explain many 
astronomical observations,
we have to develop new techniques to analyze strong gravitational lensing effects,
because the influence of the strong gravity appears when the photon 
passes the vicinity of a black hole
\cite{Cunningham:1973,Virbhadra:1999nm,Bozza:2002zj,Bozza:2008mi,Virbhadra:2008ws}.
In order to obtain definite evidence for a black hole,
the direct imaging of the shadows can be efficient~\cite{Zakharov:1994ts,Zakharov:2005ek,Schee:2008kz,Bambi:2008jg}.

To discuss such a shadow of a black hole, we have to analyze the propagation of a light ray
in the strong gravity formed by a black hole.
The Kerr solution is believed to be a unique realistic and known space-time
which well describes an astrophysical black hole formed by a gravitational collapse of a rotating body.
It is parametrized by a spin parameter $a$ and a gravitational mass $M$.
The regular horizon exists if $|a|\leq M$, and we have a Kerr black hole.
If this condition is not satisfied, the space-time has a naked  singularity.
From a theoretical standpoint, the solutions of the Einstein equations with strong
gravity generally contain a space-time singularity, as a final state of a realistic gravitational collapse.
If the event horizon appears and the space-time singularity is hidden,
a black hole is formed.
It is the so-called cosmic censorship hypothesis~\cite{Penrose}.
It implies that a naked singularity does not exist in nature.
However, this hypothesis has so far not yet been proven,
so a naked singularity is still one of the most important subjects of general relativity~\cite{Nakao:2002kc,Virbhadra:2002ju,Virbhadra:2007kw,Gyulchev:2008ff}.
There is still a possibility that the black hole candidates could be naked singularities.

The Kerr space-time will cast a variety of shadows which fully depend on the spin parameter,
the inclination angle of a black hole, and the configuration of the light emission region.
In the case that the light source is an accretion disk around
a black hole, the shadow has been intensively studied
numerically for various values of the black hole parameters and 
positions of the emission region on the disk~\cite{Falcke:1999pj,Takahashi:2004xh,Beckwith:2004ae,Broderick:2005xa,Broderick:2005at,
Takahashi:2007ac,Wu:2007bq,Huang:2007us}.
In general, the shape of the shadow depends on the parameters in a very complex way.
To extract some information about a black hole, such as a spin parameter, from such a complicated shape, 
we have to find a method by which we can determine the parameters from the observed apparent shapes. 

In this paper, we present how  to determine a spin 
parameter and an inclination angle by the direct imaging of the shadows, 
assuming that the black hole candidate is described by the Kerr space-time.
The shadows of Kerr-Newman space-times were analyzed in~\cite{Young:1976,AdeVries,Hioki:2008zw}.
It was shown that there are sensible differences between the shape of 
the shadow of a Kerr-Newman black hole and that of a naked singularity.
Here, by reanalyzing the shadows more elaborately,
we propose a method to determine those parameters 
and discuss how one can distinguish black holes from naked 
singularities for a wide range of parameters.
It may allow us to rule out the possibility by observation
that the black hole candidate is a naked singularity.

This paper is organized as follows.
In Sec.~\ref{sec:2}, we briefly summarize null-geodesics
in a Kerr space-time, and define the apparent shape of a collapsed object.
In Sec.~\ref{sec:3}, we introduce two observables and discuss
how to determine the spin parameter by observing those observables.
The summary and remarks follow in Sec.~\ref{sec:4}.
We use the geometric units, i.e., $c=G=1$ and adopt the definition
of curvatures in \cite{MTW}.

%%%%%%%%%%%%%%%%%%%%%%%%%%%%%%%%%%%%%%%%%%%%%%%%%%%%%%%%%%%%%%%%%%%%%%%
%%%%%%%%%%%%%%%%%%%%%%%%%%%%%%%%%%%%%%%%%%%%%%%%%%%%%%%%%%%%%%%%%%%%%%%
%%%%%%%%%%%%%%%%%%%%%%%%%%%%%%%%%%%%%%%%%%%%%%%%%%%%%%%%%%%%%%%%%%%%%%%
%%%%%%%%%%%%%%%%%%%%%%%%%%%%%%%%%%%%%%%%%%%%%%%%%%%%%%%%%%%%%%%%%%%%%%%
\section{Photon Orbit in Kerr Space-time and Shadow}
\label{sec:2}
%%%%%%%%%%%%%%%%%%%%%%%%%%%%%%%%%%%%%%%%%%%%%%%%%%%%%%%%%%%%%%%%%%%%%%%
%%%%%%%%%%%%%%%%%%%%%%%%%%%%%%%%%%%%%%%%%%%%%%%%%%%%%%%%%%%%%%%%%%%%%%%
%%%%%%%%%%%%%%%%%%%%%%%%%%%%%%%%%%%%%%%%%%%%%%%%%%%%%%%%%%%%%%%%%%%%%%%
%%%%%%%%%%%%%%%%%%%%%%%%%%%%%%%%%%%%%%%%%%%%%%%%%%%%%%%%%%%%%%%%%%%%%%%
\subsection{Equations of geodesic motion}
%%%%%%%%%%%%%%%%%%%%%%%%%%%%%%%%%%%%%%%%%%%%%%%%%%%%%%%%%%%%%%%%%%%%%%%
%%%%%%%%%%%%%%%%%%%%%%%%%%%%%%%%%%%%%%%%%%%%%%%%%%%%%%%%%%%%%%%%%%%%%%%
%%%%%%%%%%%%%%%%%%%%%%%%%%%%%%%%%%%%%%%%%%%%%%%%%%%%%%%%%%%%%%%%%%%%%%%
The space-time of a rotating black hole is well described by the Kerr metric  
in Boyer-Lindquist coordinates~\cite{Kerr}, which is  given by
\begin{eqnarray}
	ds^2
	&=&
	-\left(
		1-\frac{2Mr}{\rho ^2}
	\right) dt^2
	+
	\frac{\rho ^2}{\Delta} dr^2
	+
	\rho ^2 d\theta ^2
	\nonumber
	\\
	&&
	-\frac{ 4Mra\sin^2\theta }{ \rho ^2 } dt d\phi
	+
	\frac{A \sin ^2 \theta}{\rho ^2} d\phi ^2 \, ,
	\label{eq:metric}
\end{eqnarray}
 where
\begin{eqnarray}
	&&
	\Delta
	:=
	r^2 - 2Mr + a^2 \, ,
	\ \
	\rho^2
	:=
	r^2 + a^2 \cos^2 \theta \, ,
	\nonumber
	\\
	&&
	A
	:=
	\left( r^2 + a^2 \right) ^2 - \Delta a^2\sin^2\theta \, .
\end{eqnarray}
The parameters $M$ and $a$ represent,
respectively, the mass and specific angular momentum of the black hole.
The frame dragging effect is described by the off-diagonal components of the metric, $g_{t\phi}$.
The event horizon of the black hole exists if the condition $\left| a \right| \leq M$ is satisfied.
If it is not satisfied, such a space-time has a naked singularity.
In this paper we discuss not only a black hole with $\left| a \right| \leq M$
but also a naked singularity with $\left| a \right| > M$.

In the Kerr space-time, there are two Killing vector fields 
due to the assumption of stationarity and axisymmetry of the space-time.
It guarantees the existence of two conserved quantities
for a geodesic motion in the Kerr space-time
(the energy $E$ and the axial component of the angular momentum $L_z$).
We also have the Killing-Yano tensor field~\cite{Yano, Penrose},
\begin{eqnarray}
f &=& r\sin \theta d\theta \wedge \left[\left( r^2 +a^2 \right) d\phi -a dt
 \right]
\\
\nonumber
&& +a\cos \theta dr \wedge \left( dt-a\sin ^2 \theta  d\phi \right) \,,
\end{eqnarray}
which provides an additional conserved quantity (the so-called 
Carter constant ${\cal Q}$).
It makes the geodesic equation of Kerr space-time 
integrable~\cite{Carter:1968rr}.
So, introducing two conserved parameters $\xi$ and $\eta$ by
\begin{eqnarray}
        \xi=\frac{L_z}{E}~~{\rm and}~~~ \eta=\frac{{\cal Q}}{E^2} \,,	
\end{eqnarray}
we find the following null-geodesic equations:
\begin{eqnarray}
	&&
	\rho^2 \frac{dr}{d\lambda}
	=
	\pm \sqrt{{\cal R}}\, ,
	\label{eq:velocity0}
	\\
	&&
	\rho^2 \frac{d\theta}{d\lambda}
	=
	\pm \sqrt{{\it \Theta}}\, ,
	\label{eq:velocity1}
	\\
	&&
	\rho^2 \frac{dt}{d\lambda}
	=
	\frac{1}{\Delta}(A-2Mra\xi)\, ,
	\label{eq:velocity2}	
	\\
	&&
	\rho^2 \frac{d\phi}{d\lambda}
	=
	\frac{1}{\Delta} [2Mar+\xi\csc ^2\theta (\rho ^2-2Mr)]\, ,
	\label{eq:velocity3}
\end{eqnarray}
where $\lambda$ is the affine parameter, and
\begin{eqnarray}
	{\cal R}
	&:=&
        \left( r^2+a^2-a\xi \right) ^2-\Delta {\cal I}\,,
	\\
	{\it \Theta}
	&:=&
	{\cal I} -(a \sin \theta -\xi \csc \theta )^2 \,,
\label{eq:potential}
\end{eqnarray}
with 
\begin{eqnarray}
{\cal I} \left( \xi , \eta \right) := \eta +\left( a-\xi \right) ^2
\,.
\label{def_I}
\end{eqnarray}
These two conserved parameters ($\xi$ and $\eta$)
completely determine the null-geodesic~\cite{Chandrasekhar:1985kt}.
The equations for the coordinate $t$ and $\phi$
are not so important when we discuss the black hole shadow below
because of the space-time symmetries.
${\cal R}$ and ${\it \Theta}$ must be non-negative from 
Eqs.~(\ref{eq:velocity0}) and (\ref{eq:velocity1}).
This condition for ${\it \Theta}$ implies the condition such that
the pair of $\left( \xi , \eta \right)$
satisfies the constraint ${\cal I} \geq  0$.

%%%%%%%%%%%%%%%%%%%%%%%%%%%%%%%%%%%%%%%%%%%%%%%%%%%%%%%%%%%%%%%%%%%%%%%
%%%%%%%%%%%%%%%%%%%%%%%%%%%%%%%%%%%%%%%%%%%%%%%%%%%%%%%%%%%%%%%%%%%%%%%
%%%%%%%%%%%%%%%%%%%%%%%%%%%%%%%%%%%%%%%%%%%%%%%%%%%%%%%%%%%%%%%%%%%%%%%
\subsection{Apparent Shape of Collapsed Objects}
%%%%%%%%%%%%%%%%%%%%%%%%%%%%%%%%%%%%%%%%%%%%%%%%%%%%%%%%%%%%%%%%%%%%%%%
%%%%%%%%%%%%%%%%%%%%%%%%%%%%%%%%%%%%%%%%%%%%%%%%%%%%%%%%%%%%%%%%%%%%%%%
%%%%%%%%%%%%%%%%%%%%%%%%%%%%%%%%%%%%%%%%%%%%%%%%%%%%%%%%%%%%%%%%%%%%%%%
Now we investigate the shadow of a collapsed object
(a Kerr black hole or a Kerr naked singularity). 
In order to find out what such an object looks like,
we first have to define the apparent shape of a shadow.
In this paper, we assume that the light sources exist
at infinity and are distributed uniformly in all directions.
Hence the shadow is obtained by solving the
scattering problem of photons injected from any points
at infinity with any and every impact parameters.

We also assume that an observer stays at infinity ($r=+\infty$) with 
the inclination angle $i$, which is defined by
the angle  between the rotation axis of
the collapsed object and the observer's line of sight.
The celestial coordinates $(\alpha, \beta )$ of the observer are
the apparent angular distances of the image on the celestial sphere
measured from the direction of the line of sight.
$\alpha$ and $\beta$ are measured in the directions perpendicular
and parallel to  the projected rotation axis onto the celestial sphere, respectively.
 
These celestial coordinates are related to the two conserved parameters, $\xi$
and $\eta$ (and the inclination angle $i$), as
\begin{eqnarray}
&&
\alpha \left( \xi , \eta ; i \right) := \lim _{r \to \infty} 
\frac{-r p^{(\varphi )}}{p^{(t)}} = -\xi \csc i \, ,
\label{eq:celestial coordinates}
\\
&&
\beta \left( \xi , \eta ; i \right) := 
\lim _{r \to \infty}\frac{r p^{(\theta )}}{p^{(t)}} 
=\left( \eta +a^2 \cos ^2 i -\xi ^2 \cot ^2 i\right)^{1/2} \, ,
\nonumber
\end{eqnarray}
where $\left( p^{(t)}, p^{(r)}, p^{(\theta )}, p^{(\phi)} \right)$ are 
the tetrad components of the photon momentum with respect to
locally nonrotating reference frames~\cite{Bardeen:1972fi}.

Then the shadow of the collapsed object is defined as follows:
Suppose some light rays are emitted at infinity ($r=+\infty$)
and propagate near the collapsed object.
If they can reach the observer at infinity after scattering,
then its direction is not dark.
On the other hand, when they fall into the event horizon of
a black hole, the observer will never see such light rays.
Such a direction becomes dark. It makes a shadow.
We define the apparent shape of a black hole by
the boundary of the shadow~\cite{Young:1976}.
The crucial orbits are the unstable spherical
photon orbit ($r$-constant orbit), which we will define later.

In the case of a naked singularity, since there is no horizon,
almost every light ray will come back to the infinity after scattering.
However, in the Kerr space-time, there are two asymptotically flat
regions; $r\rightarrow +\infty$ and $r \rightarrow -\infty$.
Hence once the light rays cross to the region of $r<0$ and 
go away to the other infinity ($r=-\infty$), 
they will never come back to our world ($r>0$).
As a result, we find a small dark spot,
in the direction where photons escape to $r=-\infty$.
One may also find a very narrow curve, which corresponds to
the outer unstable spherical orbit (discussed later).
There is another dark region in the case of a naked singularity.
Some light rays which accidentally hit a singularity will never reach
the observer. Hence  on the celestial sphere, we will see a dark
point, in the direction where photons hit the singularity.

In order to find the apparent shape,
we first have to analyze photon orbits in a Kerr space-time.
There are two important classes of photon orbits for a shadow,
which we will discuss next.

%%%%%%%%%%%%%%%%%%%%%%%%%%%%%%%%%%%%%%%%%%%%%%%%%%%%%%%%%%%%%%%%%%%%%%%
%%%%%%%%%%%%%%%%%%%%%%%%%%%%%%%%%%%%%%%%%%%%%%%%%%%%%%%%%%%%%%%%%%%%%%%
%%%%%%%%%%%%%%%%%%%%%%%%%%%%%%%%%%%%%%%%%%%%%%%%%%%%%%%%%%%%%%%%%%%%%%%
\subsection{Important Classes of Photon Orbits }
\label{sec:2-3}
%%%%%%%%%%%%%%%%%%%%%%%%%%%%%%%%%%%%%%%%%%%%%%%%%%%%%%%%%%%%%%%%%%%%%%%
%%%%%%%%%%%%%%%%%%%%%%%%%%%%%%%%%%%%%%%%%%%%%%%%%%%%%%%%%%%%%%%%%%%%%%%
%%%%%%%%%%%%%%%%%%%%%%%%%%%%%%%%%%%%%%%%%%%%%%%%%%%%%%%%%%%%%%%%%%%%%%%
Here we discuss two important classes
of photon orbits: a spherical photon orbit  and a principal null.
%%%%%%%%%%%%%%%%%%%%%%%%%%%%%%%%%%%%%%%%%%%%%%%%%%%%%%%%%%%%%%%%%%%%%%%
%%%%%%%%%%%%%%%%%%%%%%%%%%%%%%%%%%%%%%%%%%%%%%%%%%%%%%%%%%%%%%%%%%%%%%%
%%%%%%%%%%%%%%%%%%%%%%%%%%%%%%%%%%%%%%%%%%%%%%%%%%%%%%%%%%%%%%%%%%%%%%%
%%%%%%%%%%%%%%%%%%%%%%%%%%%%%%%%%%%%%%%%%%%%%%%%%%%%%%%%%%%%%%%%%%%%%%%
%%%%%%%%%%%%%%%%%%%%%%%%%%%%%%%%%%%%%%%%%%%%%%%%%%%%%%%%%%%%%%%%%%%%%%%
\subsubsection{Spherical photon orbits} 
\label{sec:2-3-1}
%%%%%%%%%%%%%%%%%%%%%%%%%%%%%%%%%%%%%%%%%%%%%%%%%%%%%%%%%%%%%%%%%%%%%%%
%%%%%%%%%%%%%%%%%%%%%%%%%%%%%%%%%%%%%%%%%%%%%%%%%%%%%%%%%%%%%%%%%%%%%%%
%%%%%%%%%%%%%%%%%%%%%%%%%%%%%%%%%%%%%%%%%%%%%%%%%%%%%%%%%%%%%%%%%%%%%%%
One important class is the unstable spherical orbit,
which we find as follows.

If the orbit is on the equatorial plane, there is a circular orbit.
The orbit is given by the equations;
\begin{eqnarray}
\theta &=& \frac{\pi}{2}~~,
\nonumber 
\\
 {\cal R}(r) &=& 0~~,~~~
\frac{d{\cal R}}{dr}(r)  \, = \, 0 \, .
\label{eq:co}
\end{eqnarray}
For a black hole, Eq.~(\ref{eq:co}) gives
the radius of the orbit as 
\begin{eqnarray}
r&=&r_{\mathrm{circ}}^{(\pm)}:=2M\left\{
1+\cos\left[\frac{2}{3}\cos^{-1}\left(
\mp \frac{a}{M}
\right)\right]\right\} \,,
\end{eqnarray}
\begin{eqnarray}
\xi &=&\xi_{\mathrm{circ}}^{(\pm)}:=
\frac{\left[\left(r_{\mathrm{circ}}^{(\pm)}\right)^2+a^2\right]}{a}
-
\frac{2r_{\mathrm{circ}}^{(\pm)}\Delta(r_{\mathrm{circ}}^{(\pm)})}{a\left(r_{\mathrm{circ}}^{(\pm)}-M\right)}
\,,~~~
\label{sol:co}
\end{eqnarray}
where the upper sign applies to direct orbits and the lower
sign to retrograde orbits.
In this case, the other conserved parameter $\eta$ vanishes because 
the Carter constant ${\cal Q}$ is zero.

If it is a naked singularity, we have only one circular orbit,
which is a retrograde orbit with the radius
\begin{eqnarray}
r&=&\tilde r_{\mathrm{circ}}^{(-)}:=2M\left\{
1+\cosh\left[\frac{2}{3}\cosh^{-1}\left(
 \frac{a}{M}
\right)\right]\right\}
\,.
\label{sol:co2}
\end{eqnarray}

With this parameter $\xi_{\mathrm{circ}}$, the photon
can go around an infinite number of times on the circle with radius $r_{\mathrm{circ}}$.
If $|\xi|$ is slightly larger  than $|\xi_{\mathrm{circ}}^{(\pm)}|$,
the photon from infinity comes close to this circular orbit,
but goes back to infinity.
On the other hand, if $|\xi|$ is slightly smaller than 
$|\xi_{\mathrm{circ}}^{(\pm)}|$,
then the photon from infinity gets into the horizon
(or hits a ring singularity). It will never come back to our infinity ($r=+\infty$).
Hence $\xi_{\mathrm{circ}}$ gives a part of the boundary of the shadow on the equatorial plane.

For more generic photon orbits, 
the Carter constant ${\cal Q}$ does not vanish.
Such orbits are not on a two-dimensional plane, 
but turn out to be three-dimensional.
Even in that case, we can define a critical orbit 
which provides us the boundary of a shadow.
This critical orbit is the (unstable) spherical orbit.
Such an orbit is given by
\begin{eqnarray}
{\cal R}(r) = 0~~,~~~
\frac{d{\cal R}}{dr}(r)= 0 
\,,
\label{eq:so}
\end{eqnarray}
with the additional condition that there exists 
some interval 
$I (\subset [0,\pi])$ in which we find
\begin{eqnarray}
{\it \Theta}(\theta) \geq  0 ~~{\rm as}~~~ \theta \in I
\,.
\label{eq:so2}
\end{eqnarray}
The solution of Eq.~(\ref{eq:so}),   
$r=r_{\mathrm{sph}}$, gives the $r$-constant orbit.
Although the orbit is three-dimensional and could be very complicated,
it stays at the same radius.
We call it a spherical orbit.
The solutions of the spherical orbits form a one-parameter family.
Hence adopting $r_{\mathrm{sph}}$ as the ``parameter,"
we find two conserved parameters of the spherical orbits 
from Eq.~(\ref{eq:so}) as~\cite{Young:1976} 
\footnote{When we find the solution~(\ref{eq:sph}),
we assume $r^2+a^2-a\xi\neq 0$.
If $r^2+a^2-a\xi= 0$, i.e. $\xi=(r^2+a^2)/a$,
then we find ${\cal I}=0$ for ${\cal R}=0$.
On the other hand,
the condition for ${\it \Theta}\geq 0$ gives $\sin^2 \theta=\xi/a=(r^2+a^2)/a^2\geq 1$.
The possible solution is $r=0$ and $\theta=\pi/2$, which corresponds to the ring singularity.
Hence we conclude that $r^2+a^2-a\xi\neq 0$.}
\begin{eqnarray}
&&
\xi_{\mathrm{sph}}=\frac{\left[\left(r_{\mathrm{sph}}\right)^2+a^2\right]}{a}
-
\frac{2r_{\mathrm{sph}}\Delta(r_{\mathrm{sph}})}{a\left(r_{\mathrm{sph}}-M\right)} \,,
~~~~~
\label{eq:sph}
\\
&&
\eta_{\mathrm{sph}}= -\frac{r_{\mathrm{sph}}^3
\left[ r_{\mathrm{sph}}(r_{\mathrm{sph}}-3M)^2-4a^2M 
\right]}{a^2(r_{\mathrm{sph}}-M)^2}
 \, .~~~~~~~~
\label{eq:sph2}
\end{eqnarray}
The parameter $r_{\mathrm{sph}}$ is constrained by the existence condition
(\ref{eq:so2}) with Eqs.~(\ref{eq:sph}) and (\ref{eq:sph2}).

Inserting Eqs.~(\ref{eq:sph}) and (\ref{eq:sph2}) into Eq.~(\ref{def_I}), we find 
\begin{eqnarray}
{\cal I}=\frac{4r_{\mathrm{sph}}^2\Delta(r_{\mathrm{sph}})}{(r_{\mathrm{sph}}-M)^2}
\,.
\end{eqnarray}
Hence for real spherical orbits, we have to require
the condition $\Delta(r_{\mathrm{sph}})>0$, i.e.,
$r_{\mathrm{sph}}>r_+$ or $r_{\mathrm{sph}}<r_-$ for a black hole,
where $r_\pm:=M\pm \sqrt{M^2 -a^2}$ are the horizon radii,
and any radius $r_{\mathrm{sph}}$ for a naked singularity.
Since it is not a sufficient condition for the existence of a spherical orbit,
when we draw the shadow of collapsed object in \S. \ref{sec:3},
we have numerically checked whether the solution of Eq.~(\ref{eq:so}) 
satisfies the condition~(\ref{eq:so2}).

Now we analyze the stability of the spherical photon orbit.
The stability is important because an unstable spherical orbit can be
critical just as the circular orbit,
and it will provide us the boundary of the shadow.

The condition for the spherical photon orbit to be unstable is 
\begin{eqnarray}
\frac{d^2 {\cal R}}{dr^2}\left(r_{\mathrm{sph}}\right) >0
\,.
\end{eqnarray}
Assuming there exists a spherical orbit at the radius $r_{\mathrm{sph}}$,
we find that the condition for the orbit to be unstable is
\begin{eqnarray}
&&
r_{\mathrm{sph}}>r_+   ~~~
{\rm or}
~~~
r_{\mathrm{sph}} < 0
\,,
\label{eq:region_BH}
\\[.5em]
&&
~~~\mathrm{for} ~~~
 M\geq |a| ~({\rm a~black~hole)} \ ,
\nonumber
\\[1em]
&&
r_{\mathrm{sph}}>r_{0}
~~~
{\rm or}
~~~
r_{\mathrm{sph}} < 0\,, 
~~~~~
\label{eq:region_NS}
\\[.5em] 
&&
~~~\mathrm{for} ~~~
 M < |a| ~({\rm a~naked~singularity})\ ,
\nonumber
\end{eqnarray}
where 
\begin{eqnarray}
r_{0}:=M+\left[ M \left( a^2 -M^2 \right) \right] ^{1/3}.
\end{eqnarray}

For a black hole space-time, once the light rays enter into the event horizon,
they never come out.
Then the unstable spherical orbits inside the horizon
($r_{\mathrm{sph}} < 0$) do not play any role for a shadow.  
On the other hand, in the case of a naked singularity,
the inner unstable spherical orbit ($r_{\mathrm{sph}} < 0$)
is also important because some photons near this orbit may come back to the observer,
while the others may not.
It also gives a critical orbit.

Note that even if there is a stable spherical orbit,
there is no corresponding point on the celestial sphere
because any orbits near this spherical orbit never go away to infinity or come from infinity.

%%%%%%%%%%%%%%%%%%%%%%%%%%%%%%%%%%%%%%%%%%%%%%%%%%%%%%%%%%%%%%%%%%%%%%%
%%%%%%%%%%%%%%%%%%%%%%%%%%%%%%%%%%%%%%%%%%%%%%%%%%%%%%%%%%%%%%%%%%%%%%%
\subsubsection{Principal null-directions} 
\label{sec:2-3-2}
%%%%%%%%%%%%%%%%%%%%%%%%%%%%%%%%%%%%%%%%%%%%%%%%%%%%%%%%%%%%%%%%%%%%%%%
%%%%%%%%%%%%%%%%%%%%%%%%%%%%%%%%%%%%%%%%%%%%%%%%%%%%%%%%%%%%%%%%%%%%%%%
Next we consider the orbits with ${\cal I} = 0$.
Equation~(\ref{eq:potential}) implies that $\theta =\theta _0$ (constant).
This condition determines two conserved parameters as 
\begin{eqnarray}
\xi&=&\xi _\mathrm{prin}:= a \sin ^2\theta _0 \, ,
\label{eq:prin1}
\\
\eta&=&\eta _\mathrm{prin}:= -a^2 \cos ^4\theta _0 \,.
\label{eq:prin2}
\end{eqnarray}

Two tangent vectors of null-geodesics with these conserved quantities 
$\xi_\mathrm{prin}$ and $\eta_\mathrm{prin}$ are given by
\begin{eqnarray}
l^{\mu}_{\pm} \partial _\mu = \frac{r^2+a^2}{\Delta } \partial _t 
\pm  \partial _r +\frac{a}{\Delta } \partial _\phi \,,
\end{eqnarray}
which  represent the degenerate principal null-directions.
It is because they satisfy
\begin{eqnarray}
C_{abc[d}l_{\pm e]}l^b_{\pm} l^c _{\pm} =0 \,, 
\end{eqnarray}
where $C_{abcd}$ is the Weyl tensor.
The existence of such shear-free principal null-geodesics is 
guaranteed in the Petrov type D space-time.

These principal null-geodesics may give just a dark point on the celestial sphere.
Since ${\cal R}  = \rho ^4({\theta _0}) > 0$, 
the geodesics has no turning point.
The light rays go ``straight" with a constant angle $\theta_0$
from $r=+\infty$ to $r=-\infty$ (or hit on a singularity on the equatorial plane).
Hence the light ray from the direction of $\theta_0=\pi-i$
will never reach the observer.
It  constitutes a dark point.
It turns out that it is involved in the dark shadow.
Only the observer on the equatorial plane ($i=\pi/2$)
will see a dark point (see \S. \ref{sec:3-1}).

%%%%%%%%%%%%%%%%%%%%%%%%%%%%%%%%%%%%%%%%%%%%%%%%%%%%%%%%%%%%%%%%%%%%%%%
%%%%%%%%%%%%%%%%%%%%%%%%%%%%%%%%%%%%%%%%%%%%%%%%%%%%%%%%%%%%%%%%%%%%%%%
%%%%%%%%%%%%%%%%%%%%%%%%%%%%%%%%%%%%%%%%%%%%%%%%%%%%%%%%%%%%%%%%%%%%%%%
%%%%%%%%%%%%%%%%%%%%%%%%%%%%%%%%%%%%%%%%%%%%%%%%%%%%%%%%%%%%%%%%%%%%%%%
\section{Measurement of spin parameter} 
\label{sec:3}
%%%%%%%%%%%%%%%%%%%%%%%%%%%%%%%%%%%%%%%%%%%%%%%%%%%%%%%%%%%%%%%%%%%%%%%
%%%%%%%%%%%%%%%%%%%%%%%%%%%%%%%%%%%%%%%%%%%%%%%%%%%%%%%%%%%%%%%%%%%%%%%
%%%%%%%%%%%%%%%%%%%%%%%%%%%%%%%%%%%%%%%%%%%%%%%%%%%%%%%%%%%%%%%%%%%%%%%
%%%%%%%%%%%%%%%%%%%%%%%%%%%%%%%%%%%%%%%%%%%%%%%%%%%%%%%%%%%%%%%%%%%%%%%
\subsection{Expected apparent shapes}
\label{sec:3-1}
%%%%%%%%%%%%%%%%%%%%%%%%%%%%%%%%%%%%%%%%%%%%%%%%%%%%%%%%%%%%%%%%%%%%%%%
%%%%%%%%%%%%%%%%%%%%%%%%%%%%%%%%%%%%%%%%%%%%%%%%%%%%%%%%%%%%%%%%%%%%%%%
%%%%%%%%%%%%%%%%%%%%%%%%%%%%%%%%%%%%%%%%%%%%%%%%%%%%%%%%%%%%%%%%%%%%%%%
Analyzing the null-geodesics, we investigate
the shadow of a Kerr black hole or a naked singularity.
The shadow of a Kerr-Newman space-time was studied in~\cite{AdeVries, Hioki:2008zw}.
Here, we reanalyze the  shadow more elaborately 
and see whether some information about the shape and/or the size
can determine space-time parameters such as a spin.
\begin{figure}[htp]
		\begin{tabular}{ cc }
			\includegraphics[width=4.1cm]{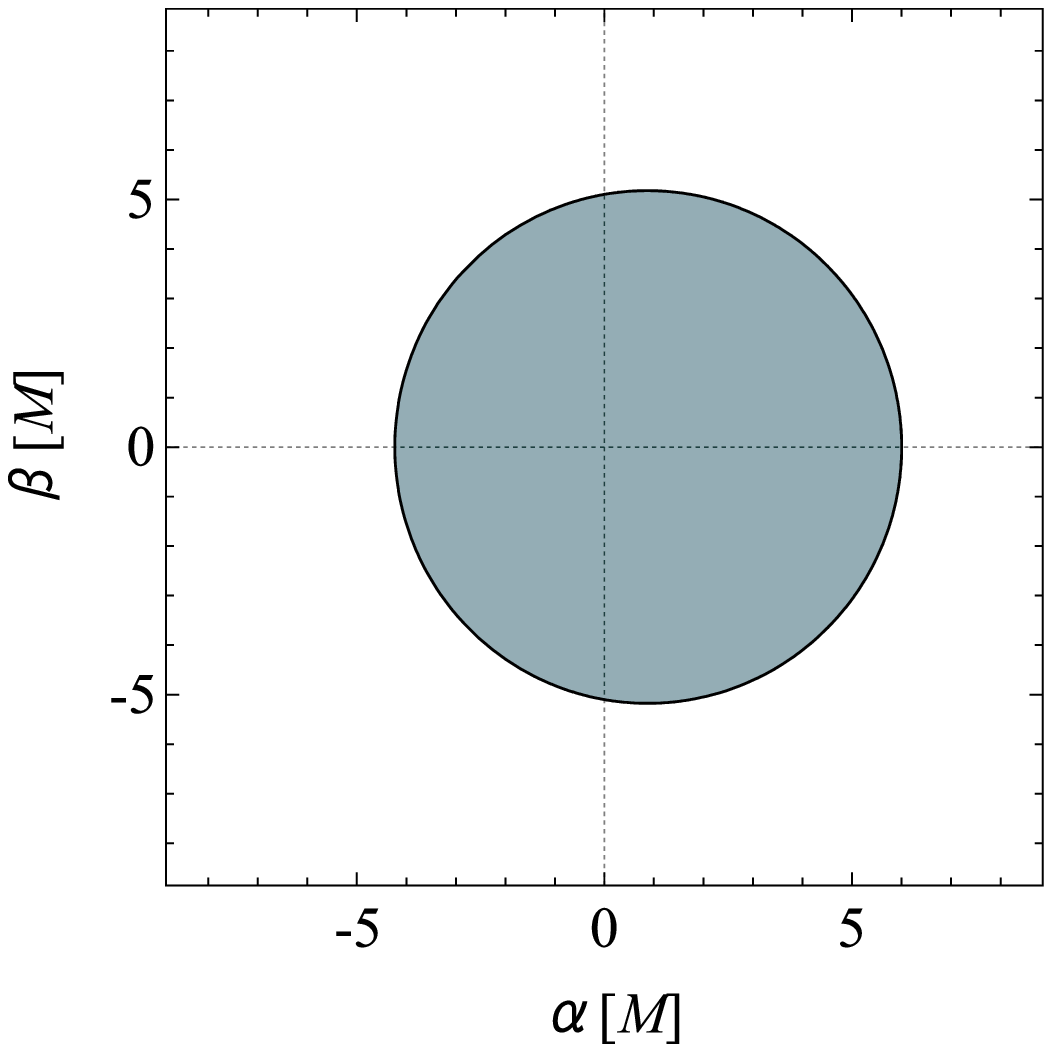} &
			\includegraphics[width=4.1cm]{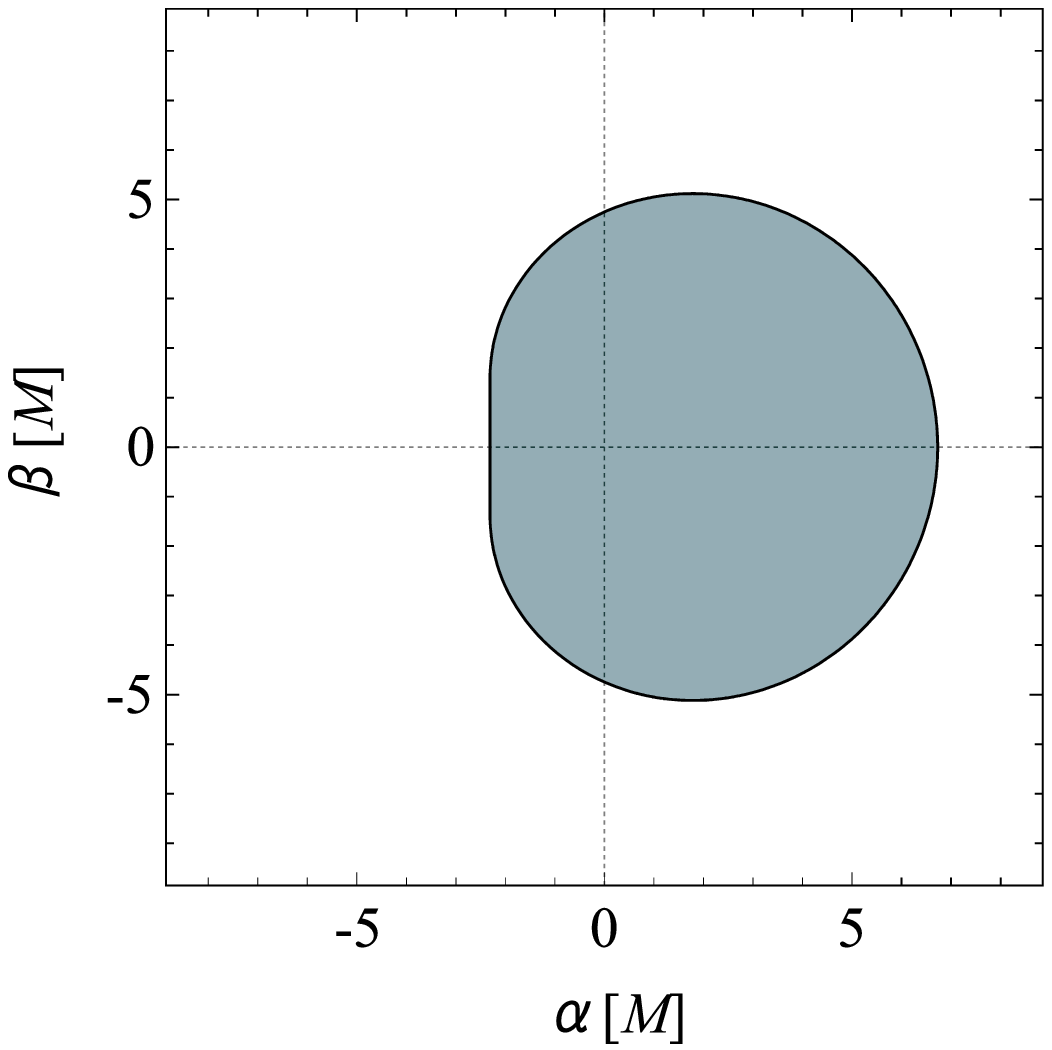} \\
			(a) $a/M=0.5$, $i=60^\circ$ &
			(b) $a/M=1$, $i=60^\circ$ \\
			\includegraphics[width=4.1cm]{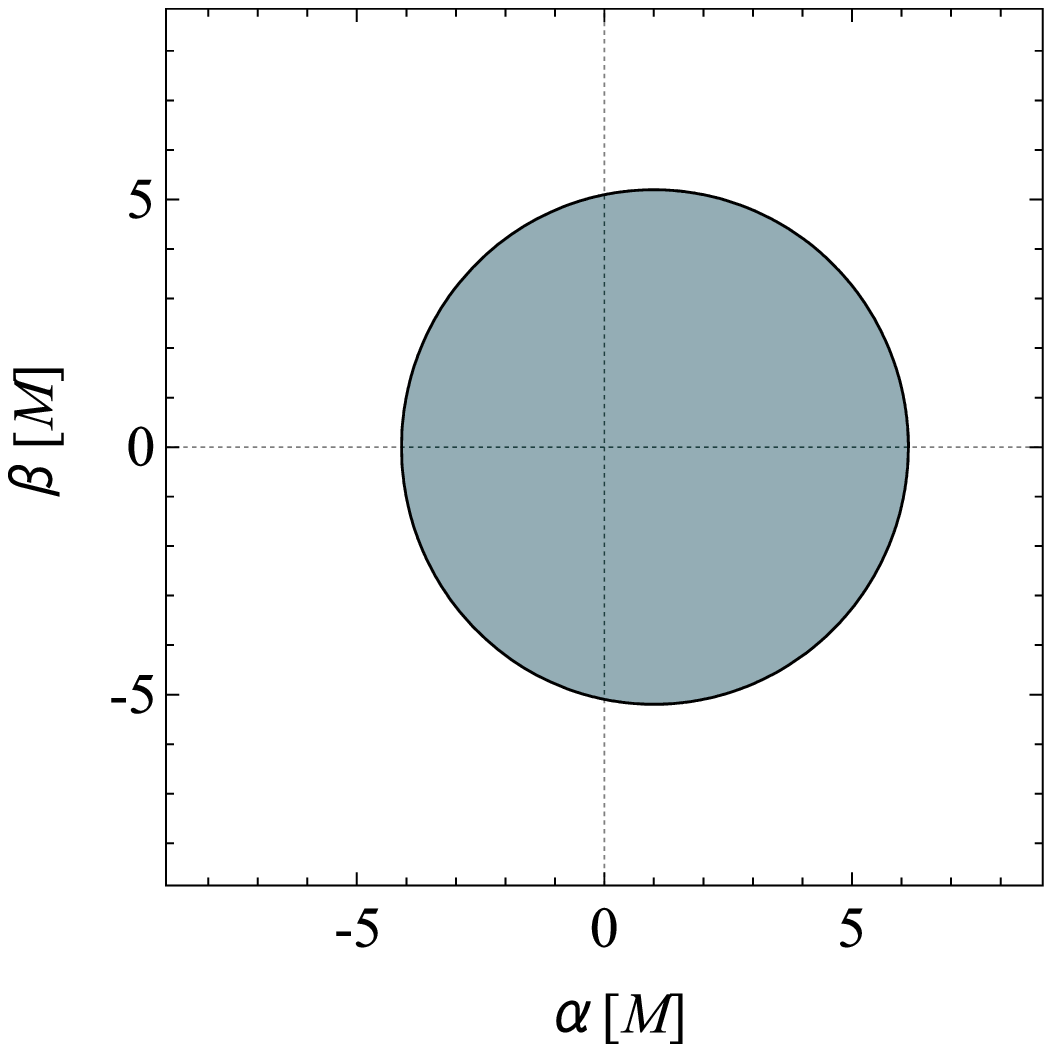} &
			\includegraphics[width=4.1cm]{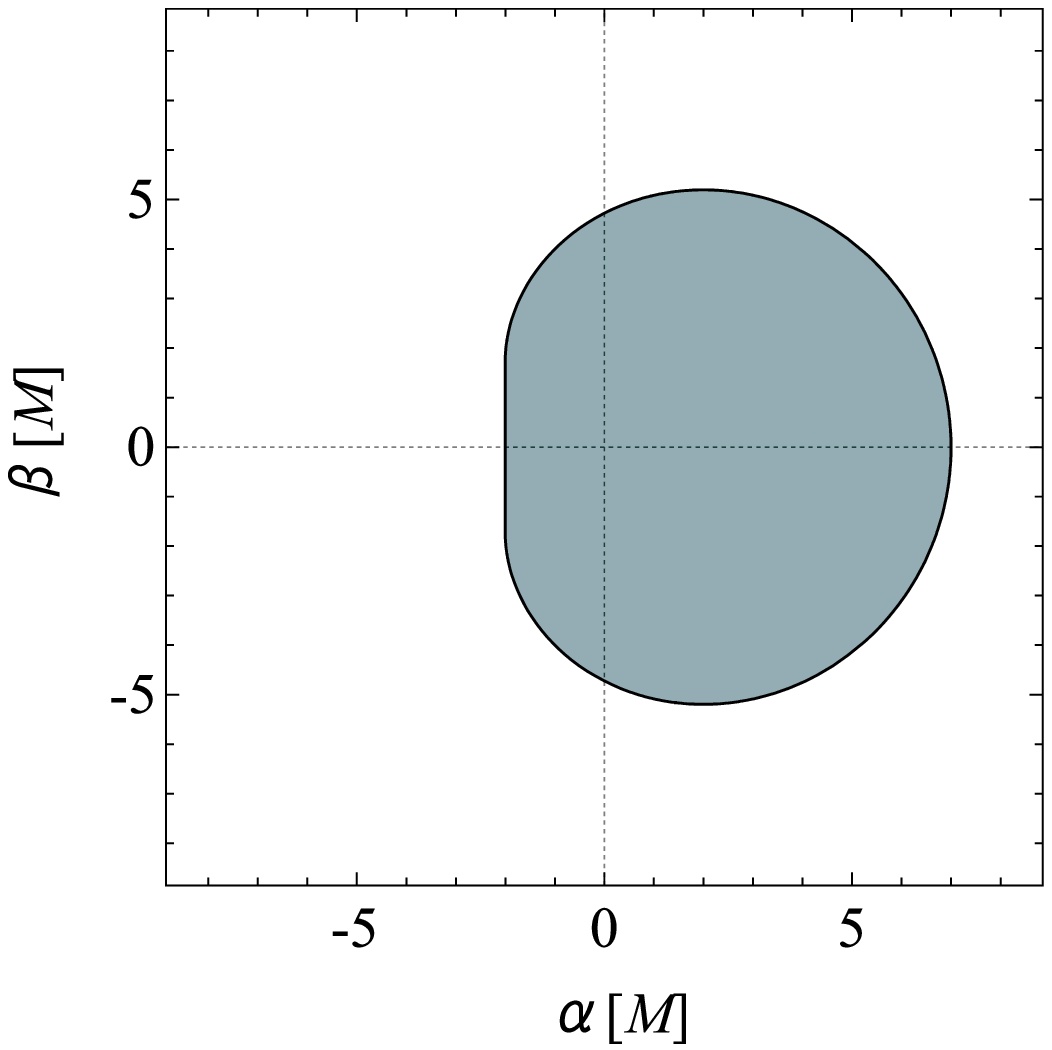} \\
			(c) $a/M=0.5$, $i=90^\circ$ &
			(d) $a/M=1$, $i=90^\circ$ \\
		\end{tabular}
\caption{\footnotesize{The shadows of Kerr black holes.
The celestial coordinates ($\alpha, \beta$) are measured in the unit of the black hole mass $M$.
} }				
		\label{fg:fig1}
\end{figure}

First we show the shadows of a Kerr black hole
using Eqs.~(\ref{eq:celestial coordinates}), (\ref{eq:sph}), 
and (\ref{eq:sph2}).
The light rays emitted at infinity will be captured by the black hole or be scattered back to infinity.
As we mentioned, the unstable spherical orbit with a positive radius 
$r_\mathrm{sph}$ gives the boundary of the shadow of a Kerr black hole.
Hence, it determines the apparent shape, which is shown in Fig.~\ref{fg:fig1}.
This shaded distorted ``disk" gives what we will see as a shadow of 
a Kerr black hole.
The inside of this distorted disk is the region where
null-geodesics is captured by the event horizon.  
If the rotation parameter $a$ is small (e.g., $a=0.5M$ as in Figs.~\ref{fg:fig1}(a) and (c)),
the shape is almost a circle,
while if it rotates very fast (e.g., $a=M$ as in Figs.~\ref{fg:fig1}(b) and (d)), the shape is distorted.
The typical feature is that the left-hand side of the disk is chipped away.
A dark point by the principal null-geodesic appears inside the disk in this case.

The distortion of the shape of the shadow 
can be understood as follows:
When a black hole has a spin,
the radius of the direct circular orbit decreases and then 
the left endpoint of the shadow moves to the right,
while the retrograde's one increases and then the right
endpoint  moves to the right as well (Fig.~\ref{fg:fig1}(c)).
However,  as the black hole rotates more rapidly,
the radius of the direct circular orbit decreases faster than 
the retrograde's one.
As the result, the left endpoint of the disk moves more to the right
compared with the right endpoint
and then the disk is distorted especially
on the left-hand side (Fig.~\ref{fg:fig1} (d)).
\begin{figure}[h]
		\begin{tabular}{ cc }
			\includegraphics[width=4.1cm]{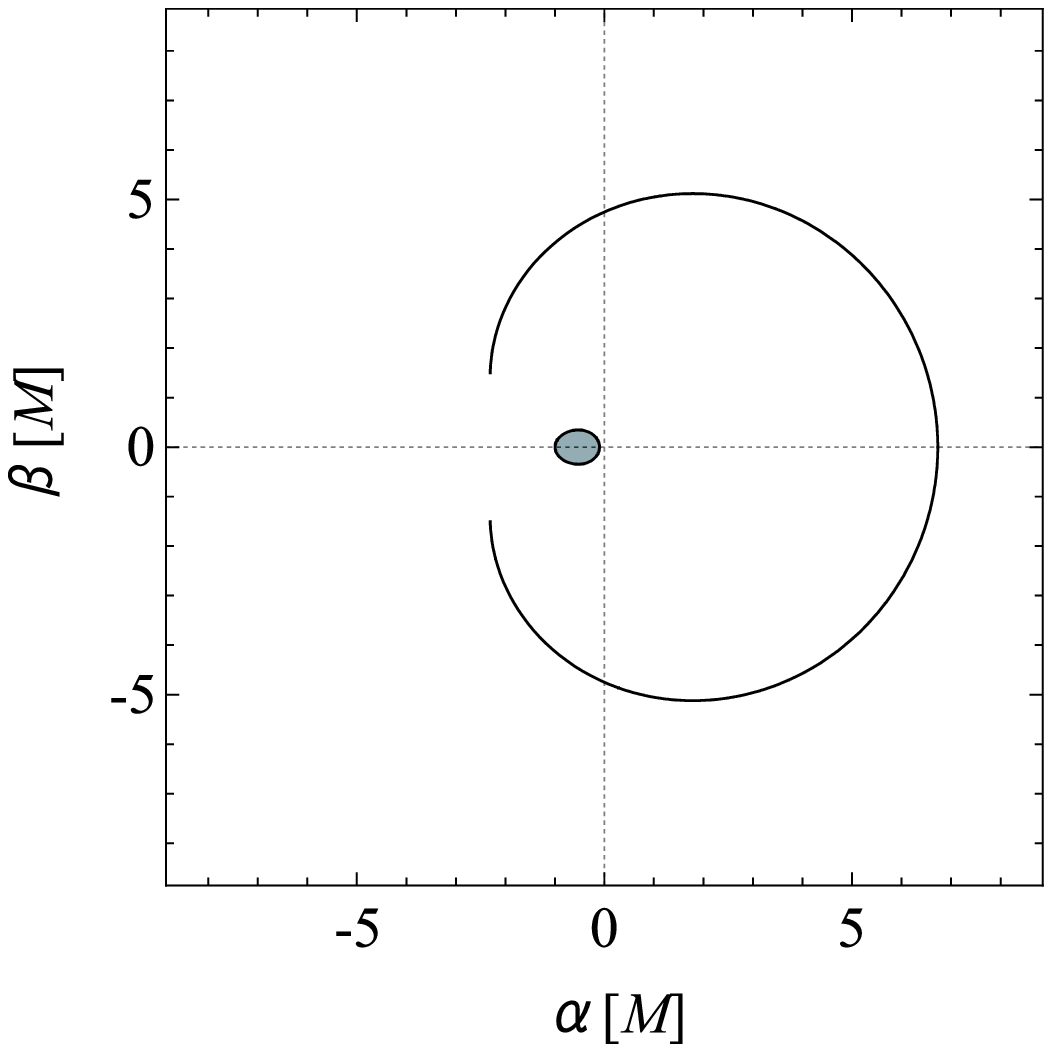} &
			\includegraphics[width=4.1cm]{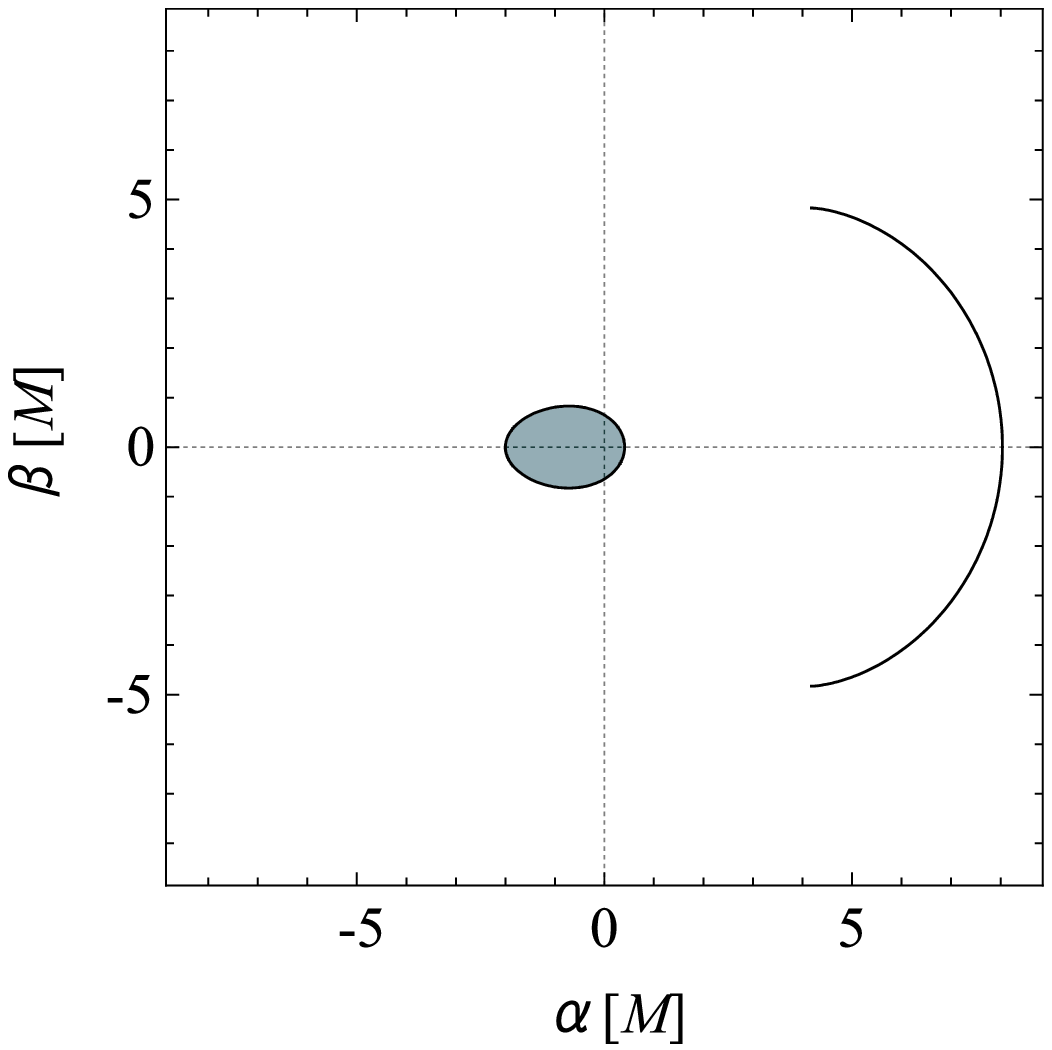} \\
			(a) $a/M=1+10^{-5}$, $i=60^\circ$ &
			(b) $a/M=2$, $i=60^\circ$ \\
			\includegraphics[width=4.1cm]{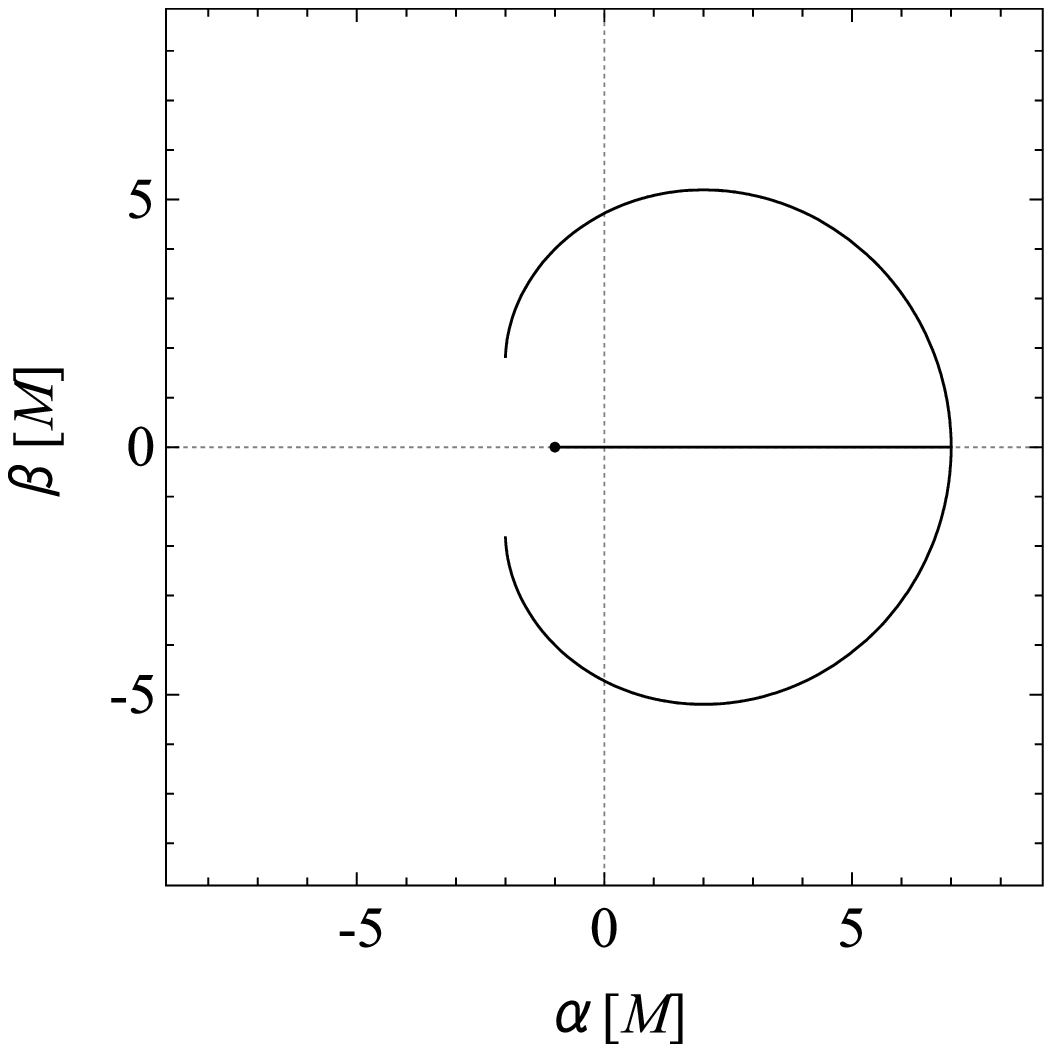} &
			\includegraphics[width=4.1cm]{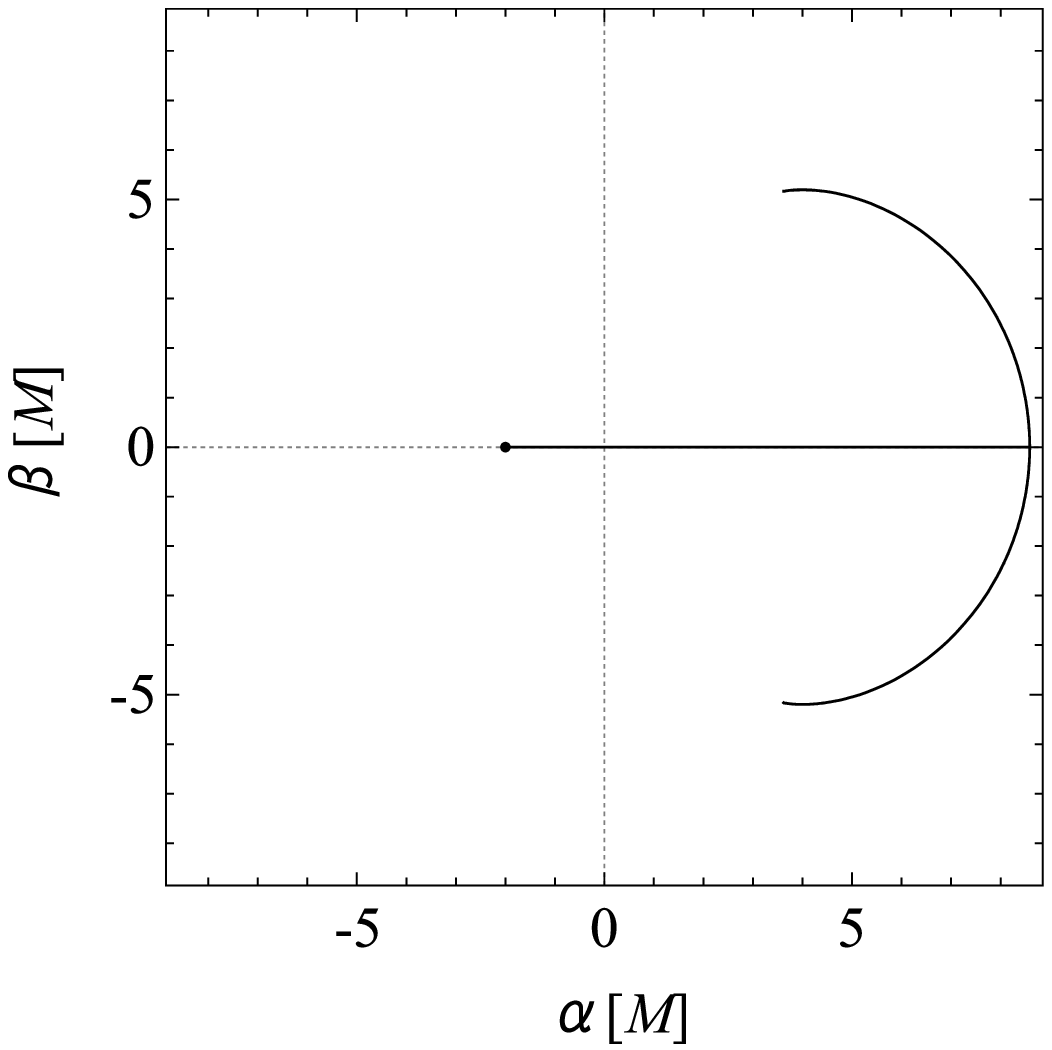} \\
			(c) $a/M=1+10^{-5}$, $i=90^\circ$ &
			(d) $a/M=2$, $i=90^\circ$ \\
		\end{tabular}
\caption{\footnotesize{The shadows of Kerr naked singularities.
 The arc (the black solid curves in (a)-(d)) is constructed
 by unstable spherical photon orbits in positive radius $r_{\mathrm{sph}}$.
 The small distorted disk (the shaded region in (a), (b)) is constructed by the null-geodesics 
which escape into the other infinity ($r=-\infty$). The straight line in (c), (d) is constructed by null-geodesics which plunge into a ring singularity. The left endpoints of those lines correspond
to the principal null-geodesics.
} }				
		\label{fg:fig2}
\end{figure}

Next we show the shadows of a Kerr naked singularity
in Fig.~\ref{fg:fig2}.
In the case of a naked singularity, the event horizon does not exist and 
then the apparent shapes drastically change from those of a black hole (see Fig.~\ref{fg:fig2}).
The unstable spherical photon orbit with a positive radius 
($r_{\mathrm{sph}}>r_{0}$)
constructs an ``arc" (the black solid curves in Figs.~\ref{fg:fig2}(a)-(d)).
This is because the photons near both sides of the arc may come back
to the observer due to the nonexistence of horizon.
While the unstable spherical photon orbits with a negative radius
($r_{\mathrm{sph}}<0$)
constructs  a dark spot 
(the small distorted disk in Figs.~\ref{fg:fig2}(a) and (b)). 
The observer will never see the light rays from such directions because they
escape into the other infinity ($r= -\infty$)
by passing through the inside of a singular ring.
It forms this dark spot.
The dark point by the principal null-geodesic appears inside the spot. 

When the observer is on the equatorial plane, 
we find that the same arc exists but
the dark spot disappears.
This is because the light rays in the direction of negative $r<0$
will always hit on a ring singularity.
Those null-geodesics construct a ``line" (the black straight line
in Figs.~\ref{fg:fig2}(c) and (d)).
Its left endpoint corresponds to the principal null-geodesics.

Changing the inclination angle from $0^\circ$ to $90^\circ$, 
the dark spot shrinks to a single point,
which corresponds to the left endpoint of the straight line
in Figs.~\ref{fg:fig2}(c) and (d), i.e., the principal null-geodesics.

%%%%%%%%%%%%%%%%%%%%%%%%%%%%%%%%%%%%%%%%%%%%%%%%%%%%%%%%%%%%%%%%%%%%%%%
%%%%%%%%%%%%%%%%%%%%%%%%%%%%%%%%%%%%%%%%%%%%%%%%%%%%%%%%%%%%%%%%%%%%%%%
%%%%%%%%%%%%%%%%%%%%%%%%%%%%%%%%%%%%%%%%%%%%%%%%%%%%%%%%%%%%%%%%%%%%%%%
\subsection{Observables}
\label{sec:3-2}
%%%%%%%%%%%%%%%%%%%%%%%%%%%%%%%%%%%%%%%%%%%%%%%%%%%%%%%%%%%%%%%%%%%%%%%
%%%%%%%%%%%%%%%%%%%%%%%%%%%%%%%%%%%%%%%%%%%%%%%%%%%%%%%%%%%%%%%%%%%%%%%
%%%%%%%%%%%%%%%%%%%%%%%%%%%%%%%%%%%%%%%%%%%%%%%%%%%%%%%%%%%%%%%%%%%%%%%
Thus far, we have seen that the parameters such as a spin parameter and 
an inclination angle determine the apparent shape of the shadow of
the Kerr space-time.
Now we study, inversely, whether it is possible to 
evaluate the spin parameter and the inclination angle 
by observing the shadow.

%%%%%%%%%%%%%%%%%%%%%%%%%%%%%%%%%%%%%%%%%%%%%%%%%%%%%%%%%%%%%%%%%%%%%%%
%%%%%%%%%%%%%%%%%%%%%%%%%%%%%%%%%%%%%%%%%%%%%%%%%%%%%%%%%%%%%%%%%%%%%%%
\subsubsection{A black hole}
%%%%%%%%%%%%%%%%%%%%%%%%%%%%%%%%%%%%%%%%%%%%%%%%%%%%%%%%%%%%%%%%%%%%%%%
%%%%%%%%%%%%%%%%%%%%%%%%%%%%%%%%%%%%%%%%%%%%%%%%%%%%%%%%%%%%%%%%%%%%%%%
\begin{figure}[ht]
		\begin{tabular}{ cc }
			\includegraphics[width=4.1cm]{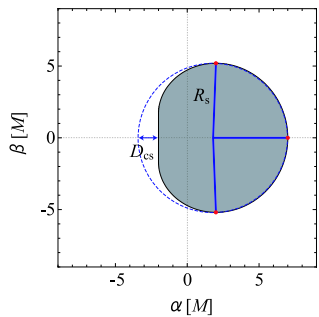}
		\end{tabular}
\caption{\footnotesize{The 
observables for the apparent shape of a Kerr
black hole are the radius $R_{\rm s}$ and 
the distortion parameter  $\delta_{\rm s}:=D_{\rm cs}/R_{\rm s}$, 
approximating it by a distorted circle,
where $D_{\rm cs}$  is the difference between  the left endpoints
of the circle and of the shadow. 
} }
		\label{fg:fig3}
\end{figure}
In the case of a Kerr black hole, we may introduce two observables which approximately
characterize the apparent shape.
First we approximate the apparent shape by a circle passing 
through three points 
which are located at the top position (A), 
the bottom position (B), and the most right end (C)
of the shadow as shown by three red points in
Fig.~\ref{fg:fig3}.
The point C corresponds to the unstable retrograde circular orbit
when seen from an observer on the equatorial plane.
We define the radius $R_{\rm s}$ of the shadow by the radius of 
this approximated circle.
We also take into account the dent in the left-hand side of the shadow
(see Figs.~\ref{fg:fig1}(b) and (d)). 
The size of this dent is evaluated by $D_{\rm cs}$,
which is the difference between  the left endpoints
of the circle and of the shadow (see Fig.~\ref{fg:fig3}).
Then we define the distortion parameter $\delta_{\rm s}$
of the shadow by $\delta_{\rm s} := D_{\rm cs} /R_{\rm s}$.
Thus we adopt these two variables ($R_{\rm s}$ and $\delta_{\rm s}$) 
as observables in astronomical observation.
\begin{figure}[t]
		\begin{tabular}{ c }
			\includegraphics[width=6.1cm]{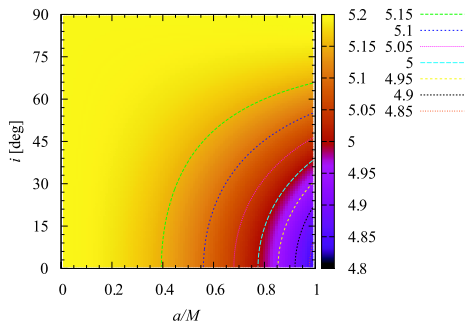} \\ 
			(a)	The radius of the shadow $R_{\rm s}$ 
			~~~~~~~~~~~~~~~~~\\[1em]
			\includegraphics[width=6cm]{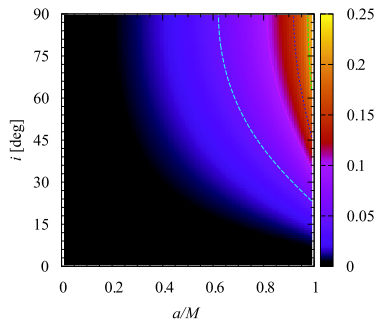} \\
			(b) The distortion parameter
                        of the shadow $\delta_{\rm s}$
		\end{tabular}
\caption{\footnotesize{(a) The contour map of
radii of the shadows of a Kerr black hole. (b) The contour map of
their distortion parameters.
} }
		\label{fg:fig4} 
\end{figure}

We show the contour maps of the
radius $R_{\rm s}$ and of the distortion parameter $\delta_{\rm s}$ in terms of
a spin parameter $a$ and an inclination angle $i$ in Fig.~\ref{fg:fig4}.
One can see from Fig.~\ref{fg:fig4}(a) that the radius decreases as the spin 
parameter becomes larger but the inclination angle gets smaller.
This is because of the following reason:
The horizon radius decreases as $a$ increases, and then the radius of
 the shadow also decreases when we observe near the rotation axis ($i\approx 0^\circ$).
When we observe near the equatorial plane ($i\approx 90^\circ$), however,
the radius of the shadow is insensitive to the rotation parameter $a$ because 
$r_{\rm circ}^{(-)}$ gets large due to the frame dragging effect.

If one observes $R_{\rm s}$ as well as the black hole mass $M$ 
and the inclination angle $i$,
one can determine the spin parameter from this figure.
On the other hand, from Fig.~\ref{fg:fig4}(b),
we find the different tendency for the distortion
of the apparent shape, i.e., 
$\delta_{\rm s}$ increases as the 
spin parameter gets larger as well as the inclination angle increases.
This is because we find the large distortion for the observer near
the equatorial plane ($i\approx 90^{\circ}$) 
due to the frame dragging effect, but
no distortion appears for the observer on the rotation axis
($i\approx 0^{\circ}$).

\begin{figure}[t]
		\begin{tabular}{ cc }
			\includegraphics[width=5.1cm]{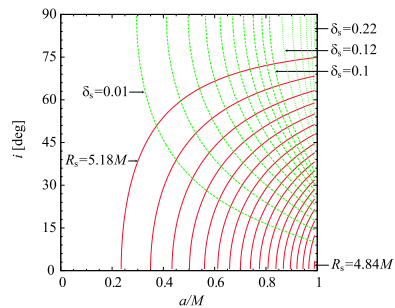}\\
		\end{tabular}
\caption{\footnotesize{The contour maps of the radius of the arc $R_{\rm s}$
(the red solid curves) and the distorted parameter $\delta_{\rm s}$ (the green dashed curves). 
The contours are those for $R_{\rm s}/M=4.84, \cdots, 5.18$,
with the contour interval being $0.02$, 
and $\delta_{\rm s}=0.01, \cdots, 0.1$ and $0.12, \cdots, 0.22$,
with the contour intervals being $0.01$  and $0.02$, respectively.
We can evaluate the spin parameter $a$ of the Kerr black hole
and the inclination angle $i$ of the observer. 
} }
		\label{fg:fig5} 
\end{figure}

If one observes $\delta_{\rm s}$ as well as $M$ and $i$,
one can also determine the spin parameter.
However, it may be very difficult to determine the inclination angle $i$.
Therefore we shall combine two contour maps for those observables
($R_{\rm s}$ and $\delta_{\rm s}$).
The one-to-one correspondence between ($a$ and $i$)
and ($R_{\rm s}$ and $\delta_{\rm s}$) is very clear as shown in 
 Fig.~\ref{fg:fig5}.
Hence if one measures the radius $R_{\rm s}$ and the distortion parameter
$\delta_{\rm s}$ by observation,
the spin parameter $a$ and the inclination angle $i$ could be 
determined by use of Fig.~\ref{fg:fig5}.
For example, assuming that we know a black hole mass $M$,
if we find two observables as $R_{\rm s}=5.1 M$ and  $\delta_{\rm s}=0.05$,
we can conclude that $a=0.784M$ and $i=44.1^\circ$.
So we may use this method to search for the parameters, $a$ and $i$. 

%%%%%%%%%%%%%%%%%%%%%%%%%%%%%%%%%%%%%%%%%%%%%%%%%%%%%%%%%%%%%%%%%%%%%%%
%%%%%%%%%%%%%%%%%%%%%%%%%%%%%%%%%%%%%%%%%%%%%%%%%%%%%%%%%%%%%%%%%%%%%%%
\subsubsection{A naked singularity}
%%%%%%%%%%%%%%%%%%%%%%%%%%%%%%%%%%%%%%%%%%%%%%%%%%%%%%%%%%%%%%%%%%%%%%%
%%%%%%%%%%%%%%%%%%%%%%%%%%%%%%%%%%%%%%%%%%%%%%%%%%%%%%%%%%%%%%%%%%%%%%%
\begin{figure}[ht]
		\begin{tabular}{ cc }
			\includegraphics[width=4.1cm]{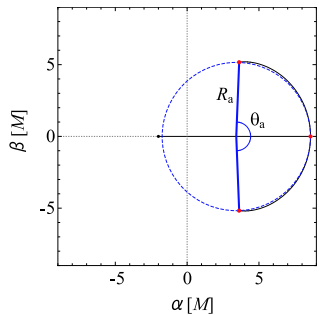} \\
		\end{tabular}
\caption{\footnotesize{The observables for the apparent shape of
a Kerr naked singularity are the radius $R_{\rm a}$ and
the central angle $\vartheta_{\rm a}$, approximating it by an arc. 
} }
		\label{fg:fig6} 
\end{figure}
In the case of a Kerr naked singularity, the shadow 
consists of two parts (the arc and 
the dark spot [or the straight line]).
One interesting shape is the arc,
which may not be observable because it is one-dimensional and then its measure is zero. 
In realistic observation, however,
the neighborhood of the arc will also be darkened to be observed as a dark ``lunate" shadow.
If it is the case, we have a chance to observe a shadow of a naked singularity.
We approximate this dark lunate shadow by the arc with the radius 
$R_{\rm a}$ and the central angle $\vartheta_{\rm a}$ as Fig.~\ref{fg:fig6}.
\begin{figure}[t]
		\begin{tabular}{ c }
			\includegraphics[width=5.9cm]{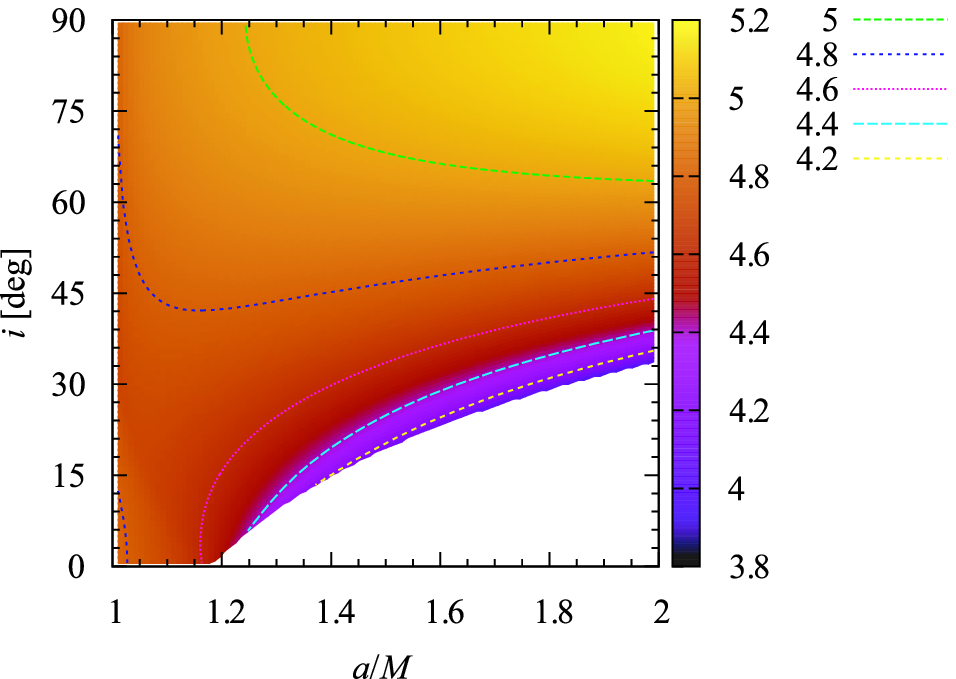} \\ 
			(a) The radius of the ``arc" $R_{\rm a}$ 
			~~~~~~~\\[1em]
			\includegraphics[width=6cm]{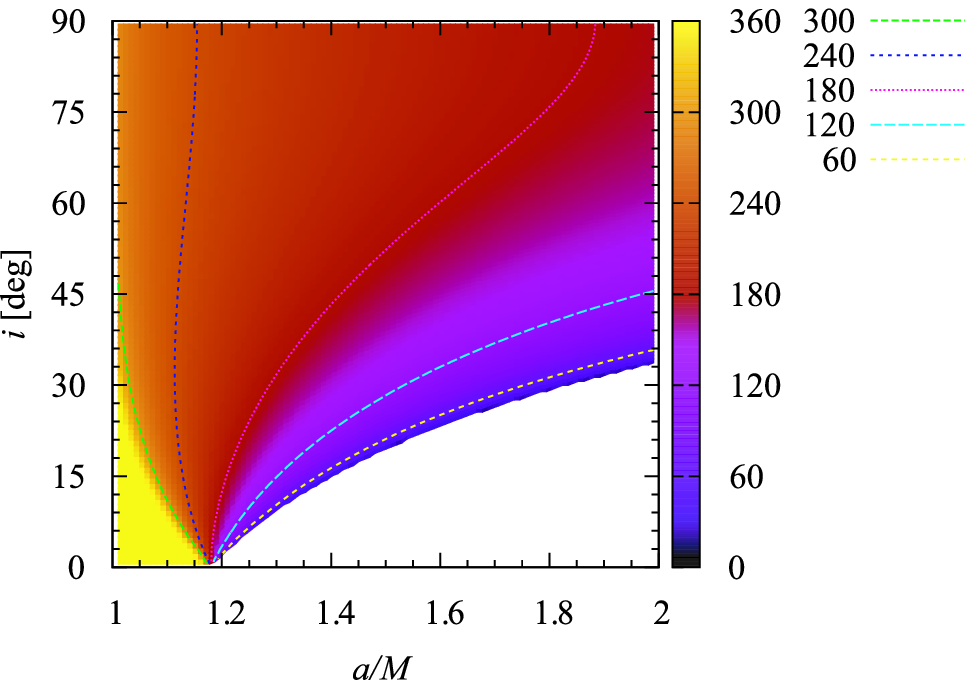} \\
			(b) The central angle  of the ``arc" 
                        $\vartheta_{\rm a}$
		\end{tabular}
\caption{\footnotesize{(a) The contour map of radii of the arcs of a Kerr naked singularity. 
(b) The contour map of central angles of the arcs.
In the blank spaces (white areas) in (a) and (b), the arc does not exist.
} }
		\label{fg:fig7}
\end{figure}

We show the contour maps of the radius  of the arc  $R_{\rm a}$ and 
of its central angle $\vartheta_{\rm a}$ in Fig.~\ref{fg:fig7}.
As we see in Figs.~\ref{fg:fig7}(a) and (b),
there is a blank space (a white area) in the right-bottom corner,
in which the arc does not appear.
It is because there exists no unstable spherical orbit
with a positive radius in this area.
From Fig.~\ref{fg:fig7}(a), we find that the radius increases as the 
inclination angle gets larger if 
the spin parameter is larger than $a\sim 1.1M$.
This behavior is similar to the case of the radius of the black hole
shadow (Fig. \ref{fg:fig4} (a)), except for the blank space.
\begin{figure}[tp]
		\begin{tabular}{ cc }
	\includegraphics[width=4.7cm]{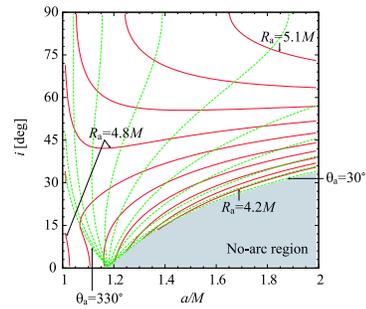} \\
		\end{tabular}
\caption{\footnotesize{The contour map  of the radius of the arc $R_{\rm a}$ (the red solid curves)
and its central angle $\vartheta_{\rm a}$ (the green dashed curves). 
The contours are those for $R_{\rm a}/M=4.2, \cdots, 5.1$,
with the contour interval being $0.1$
and $\vartheta_{\rm a}=30^\circ, \cdots, 330^\circ$,
with the contour interval being $30^\circ$.
If one can observe $\vartheta_{\rm a}$ and $R_{\rm a}$,
the spin parameter $a$ of the Kerr naked singularity and the inclination angle $i$ could be evaluated,
although there exists a degenerate region near the blank space shaded in gray.
} }
		\label{fg:fig8} 
\end{figure}

The radius is finite in the neighborhood of the blank space,
and vanishes incontinuously on the boundary.
From Fig.~\ref{fg:fig7}(b), the central angle approaches $360^\circ $ when
both a spin parameter and an inclination angle are sufficiently small.
That is to say, the arc closes and becomes a distorted ``circle" 
in such parameter region.
The central angle shows a steep decline to $0^\circ$ 
in the neighborhood of the blank space.
On the boundary, the arc vanishes with a finite  radius.

In Fig.~\ref{fg:fig8}, we show the contour maps of the radius and of the central angle
in terms of the spin parameter and inclination angle.
Since these contours also give one-to-one correspondence
between ($R_{\rm a}$ and $\vartheta_{\rm a}$) and ($a$ and $i$), 
we may be able to evaluate a spin parameter and an inclination angle
by observing those two observables $R_{\rm a}$ and $\vartheta_{\rm a}$, 
just as the same as the case of a black hole.

When the inclination is larger than $45^\circ$ (the upper half part of Fig.~\ref{fg:fig8}),
one may easily determine the parameters $a$ and $i$
from the observation of $R_{\rm a}$ and $\vartheta_{\rm a}$.
However, near the blank space, the contours of 
the spin parameter and the inclination angle become degenerate.
One may find that it is hard to evaluate those parameters
by real astronomical observations.

\begin{figure}[b]
		\begin{tabular}{ c }
	\includegraphics[width=4.1cm]{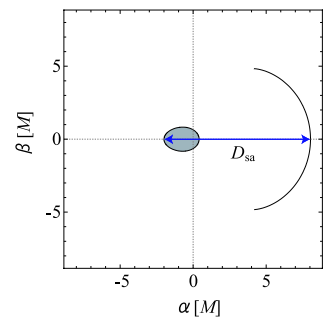} \\
				\end{tabular}
\caption{\footnotesize{
The separation distance between the left endpoint of the dark spot
and the center of the arc, which we denote $D_{\rm sa}$.
We then define the aspect ratio of the shadow of a Kerr naked singularity;
$\mathcal{A} _{\rm sa} :=D_{\rm sa} /R_{\rm a}$, where $R_{\rm a}$ is the arc radius.}}
		\label{fg:fig9}
\end{figure}
\begin{figure}[h]
		\begin{tabular}{ c }
	\includegraphics[width=5.8cm]{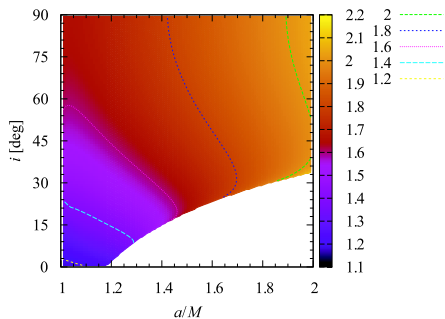} \\ 
		\end{tabular}
\caption{\footnotesize{The contour map of the aspect ratio 
of the shadow of a Kerr naked singularity,
$\mathcal{A}_{\rm sa}$.}}
\label{fg:fig10}
\end{figure}
\begin{figure}[t]
		\begin{tabular}{ cc }
	\includegraphics[width=4.7cm]{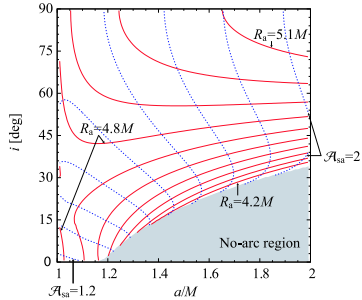} \\
		\end{tabular}
\caption{\footnotesize{The contour maps of the aspect ratio 
$\mathcal{A}_{\rm sa}$ (the blue dashed curves) and of the radius of the 
arc $R_{\rm a}$ (the red solid curves).
The contours are those for $R_{\rm a}/M=4.2, \cdots, 5.1$,
with the contour interval being $0.1$ and $\mathcal{A}_{\rm sa}=1.2, \cdots, 2.0$,
with the contour interval being $0.1$.
The one-to-one correspondence between ($R_{\rm a}$ and $\mathcal{A}_{\rm sa}$)
and ($a$ and $i$) is very clear.
We can use this contour map to determine the spin parameter and the inclination angle.
} }
		\label{fg:fig11}
\end{figure}

However, in the case of a naked singularity, we have another part of 
the shadow, i.e., the dark spot (or the dark line).
We can introduce another observable, 
the aspect ratio $\mathcal{A}_{\rm sa}:=D_\mathrm{sa}/R_\mathrm{a}$
by defining the separation distance between the left endpoint of the dark spot
 (or the left endpoint of the line) and the center of the arc, which we 
denote $D_{\rm sa}$ (see Fig.~\ref{fg:fig9}).
We call $\mathcal{A}_{\rm sa}$ an aspect ratio since it is the 
ratio of the horizontal size of the shadow to the vertical one as a whole.
We show the contour map of this observable in Fig.~\ref{fg:fig10}.
This behavior is similar to the case of the deformation parameter
of the black hole shadow (Fig. \ref{fg:fig4} (b)), except for the blank space.
From Fig.~\ref{fg:fig10}, we find this aspect ratio $\mathcal{A}_{\rm sa}$
highly depends on $a$, which is very different from $R_{\rm a}$.
Hence a pair of $R_{\rm a}$ and $\mathcal{A}_{\rm sa}$ may be better to use
for the determination of $a$ and $i$.
In fact, combining the two contour maps of $R_{\rm a}$ and $\mathcal{A}_{\rm sa}$,
we find a clear one-to-one correspondence as shown in Fig.~\ref{fg:fig11}.
Hence if we observe both $R_{\rm a}$ and $\mathcal{A}_{\rm sa}$, we can easily
determine $a$ and $i$.

Note that we may find one-to-two correspondence for some parameters, e.g.,
$R_{\rm a}/M=4.8$ and $\mathcal{A}_{\rm sa}=1.6$.
Even in that case, such information
is still very useful because we have only two choices, e.g., 
$a/M=1.21$ or $1.02$ for the above data.

%%%%%%%%%%%%%%%%%%%%%%%%%%%%%%%%%%%%%%%%%%%%%%%%%%%%%%%%%%%%%%%%%%%%%%%
%%%%%%%%%%%%%%%%%%%%%%%%%%%%%%%%%%%%%%%%%%%%%%%%%%%%%%%%%%%%%%%%%%%%%%%

%%%%%%%%%%%%%%%%%%%%%%%%%%%%%%%%%%%%%%%%%%%%%%%%%%%%%%%%%%%%%%%%%%%%%%%
%%%%%%%%%%%%%%%%%%%%%%%%%%%%%%%%%%%%%%%%%%%%%%%%%%%%%%%%%%%%%%%%%%%%%%%
%%%%%%%%%%%%%%%%%%%%%%%%%%%%%%%%%%%%%%%%%%%%%%%%%%%%%%%%%%%%%%%%%%%%%%%
%%%%%%%%%%%%%%%%%%%%%%%%%%%%%%%%%%%%%%%%%%%%%%%%%%%%%%%%%%%%%%%%%%%%%%%
\section{Conclusion} 
\label{sec:4}
%%%%%%%%%%%%%%%%%%%%%%%%%%%%%%%%%%%%%%%%%%%%%%%%%%%%%%%%%%%%%%%%%%%%%%%
%%%%%%%%%%%%%%%%%%%%%%%%%%%%%%%%%%%%%%%%%%%%%%%%%%%%%%%%%%%%%%%%%%%%%%%
A black hole casts a shadow as an optical 
appearance because of its strong gravitational field.
The apparent shape of the shadow is distorted 
mainly by its spin parameter and the inclination angle.
In this paper, we investigated whether it is possible to determine the spin
parameter and the inclination angle by observing the apparent shape 
of the shadow of a compact object, which is assumed to be described
by Kerr space-time.
Introducing two observables, the radius $R_{\rm s}$ and 
the distortion parameter $\delta_{\rm s}$, characterizing the 
apparent shape,
we found that the spin parameter and the inclination angle
of a Kerr black hole can be determined by measuring those 
observables.
It will rule out the possibility that a black hole candidate
is a naked singularity, if we observe them.
We have also extended this technique to the case of a Kerr naked singularity.
The result may provide us with a practical method for future advanced interferometers.

Of course, we need further analysis to apply the present approach 
to realistic observations.
First we  have to evaluate the luminous flux in each direction.
Then we can predict how dark in the directions around the shadow.
It is especially important in the case of a naked singularity,
because we have a one-dimensional dark arc as the shadow.

We usually expect that the light source is
at a finite distance and the size is also finite such as an accretion disk.
In that case, we may find shapes of shadows different from the present one.
We should study several realistic light source models 
(a compact object with an accretion disk, or a gravitationally collapsing object, etc)
in order to apply the present method. 
However, note that the present shape gives the region in which direction
we can never observe any light from any light sources.
Hence, even if we have more realistic light source,
once we can find a part of the present shape,
we may give some limits on the spin parameter and inclination angle.

%%%%%%%%%%%%%%%%%%%%%%%%%%%%%%%%%%%%%%%%%%%%%%%%%%%%%%%%%%%%%%%%%%%%%%%
%%%%%%%%%%%%%%%%%%%%%%%%%%%%%%%%%%%%%%%%%%%%%%%%%%%%%%%%%%%%%%%%%%%%%%%
%%%%%%%%%%%%%%%%%%%%%%%%%%%%%%%%%%%%%%%%%%%%%%%%%%%%%%%%%%%%%%%%%%%%%%%
\section*{Acknowledgements}
%%%%%%%%%%%%%%%%%%%%%%%%%%%%%%%%%%%%%%%%%%%%%%%%%%%%%%%%%%%%%%%%%%%%%%%
%%%%%%%%%%%%%%%%%%%%%%%%%%%%%%%%%%%%%%%%%%%%%%%%%%%%%%%%%%%%%%%%%%%%%%%
This work was supported in part by the Grant-in-Aid for
Scientific Research Fund of the JSPS Nos. 19540308 and 19$\cdot$07795,
and by the Japan-U.K. Research Cooperative Program.

%%%%%%%%%%%%%%%%%%%%%%%%%%%%%%%%%%%%%%%%%%%%%%%%%%%%%%%%%%%%%%%%%%%%%%%
%%%%%%%%%%%%%%%%%%%%%%%%%%%%%%%%%%%%%%%%%%%%%%%%%%%%%%%%%%%%%%%%%%%%%%%
%%%%%%%%%%%%%%%%%%%%%%%%%%%%%%%%%%%%%%%%%%%%%%%%%%%%%%%%%%%%%%%%%%%%%%%
%%%%%%%%%%%%%%%%%%%%%%%%%%%%%%%%%%%%%%%%%%%%%%%%%%%%%%%%%%%%%%%%%%%%%%%

%%%%%%%%%%%%%%%%%%%%%%%%%%%%%%%%%%%%%%%%%%%%%%%%%%%%%%%%%%%%%%%%%%%%%%%
%%%%%%%%%%%%%%%%%%%%%%%%%%%%%%%%%%%%%%%%%%%%%%%%%%%%%%%%%%%%%%%%%%%%%%%
%%%%%%%%%%%%%%%%%%%%%%%%%%%%%%%%%%%%%%%%%%%%%%%%%%%%%%%%%%%%%%%%%%%%%%%
%%%%%%%%%%%%%%%%%%%%%%%%%%%%%%%%%%%%%%%%%%%%%%%%%%%%%%%%%%%%%%%%%%%%%%%


\begin{thebibliography}{99}

%\cite{Eisenhauer:2005cv}
\bibitem{Eisenhauer:2005cv}
  F.~Eisenhauer {\it et al.},
  %``SINFONI in the Galactic Center: young stars and IR flares in the central
  %light month,''
  Astrophys.\ J.\  {\bf 628}, 246 (2005)
  [arXiv:astro-ph/0502129].
  %%CITATION = ASJOA,628,246;%%

%\cite{Tanaka:1995en}
\bibitem{Tanaka:1995en}
  Y.~Tanaka {\it et al.},
  %``Gravitationally Redshifted Emission Implying an Accretion Disk and Massive
  %Black Hole in the Active Galaxy MCG:-6-30-15,''
  Nature {\bf 375}, 659 (1995).
  %%CITATION = NATUA,375,659;%%
  
%\cite{Reynolds:2002np}
\bibitem{Reynolds:2002np}
  C.~S.~Reynolds and M.~A.~Nowak,
  %``Fluorescent iron lines as a probe of astrophysical black hole systems,''
  Phys.\ Rept.\  {\bf 377}, 389 (2003)
  [arXiv:astro-ph/0212065].
  %%CITATION = PRPLC,377,389;%%

%\cite{Cash:2000exp}
\bibitem{Cash:2000exp}
  W.~Cash, A.~Shipley, S.~Osterman and M.~Joy,
  %``Laboratory detection of X-ray fringes with a grazing-incidence interferometer,''
  Nature {\bf 407}, 160 (2000).

%\cite{Hirabayashi:2005kc}
\bibitem{Hirabayashi:2005kc}
  H.~Hirabayashi {\it et al.},
  %``On the Near-term Space VLBI Mission VSOP-2,''
  arXiv:astro-ph/0501020.
  %%CITATION = ASTRO-PH/0501020;%%

%\cite{Doeleman:2008qh}
\bibitem{Doeleman:2008qh}
  S.~Doeleman {\it et al.},
  %``Event-horizon-scale structure in the supermassive black hole candidate at
  %the Galactic Centre,''
  Nature {\bf 455}, 78 (2008)
  [arXiv:0809.2442 [astro-ph]].
  %%CITATION = NATUA,455,78;%%

%\cite{Doeleman:2008xq}
\bibitem{Doeleman:2008xq}
  S.~S.~Doeleman, V.~L.~Fish, A.~E.~Broderick, A.~Loeb and A.~E.~E.~Rogers,
  %``Methods for detecting flaring structures in Sagittarius A* with high
  %frequency VLBI,''
  Astrophys.\ J.\  {\bf 695}, 59 (2009)
  [arXiv:0809.3424 [astro-ph]].
  %%CITATION = ASJOA,695,59;%%

%\cite{Cunningham:1973}
\bibitem{Cunningham:1973}
   J.~M.~Cunningham and C.~T.~Bardeen,
   %``The Optical Appearance of a Star Orbiting an Extreme Kerr Black Hole,''
   Astrophys.\ J.\  {\bf 183}, 237 (1973)

%\cite{Virbhadra:1999nm}
\bibitem{Virbhadra:1999nm}
  K.~S.~Virbhadra and G.~F.~R.~Ellis,
  %``Schwarzschild black hole lensing,''
  Phys.\ Rev.\  D {\bf 62} (2000) 084003
  [arXiv:astro-ph/9904193].
  %%CITATION = PHRVA,D62,084003;%%
  
%\cite{Bozza:2002zj}
\bibitem{Bozza:2002zj}
  V.~Bozza,
  %``Gravitational lensing in the strong field limit,''
  Phys.\ Rev.\  D {\bf 66}, 103001 (2002)
  [arXiv:gr-qc/0208075].
  %%CITATION = PHRVA,D66,103001;%%

%\cite{Bozza:2008mi}
\bibitem{Bozza:2008mi}
  V.~Bozza,
  %``Optical caustics of Kerr spacetime: the full structure,''
  Phys.\ Rev.\  D {\bf 78}, 063014 (2008)
  [arXiv:0806.4102 [gr-qc]].
  %%CITATION = PHRVA,D78,063014;%%

%\cite{Virbhadra:2008ws}
\bibitem{Virbhadra:2008ws}
  K.~S.~Virbhadra,
  %``Relativistic images of Schwarzschild black hole lensing,''
  Phys.\ Rev.\  D {\bf 79}, 083004 (2009)
  [arXiv:0810.2109 [gr-qc]].
  %%CITATION = PHRVA,D79,083004;%%

%\cite{Zakharov:1994ts}
\bibitem{Zakharov:1994ts}
  A.~F.~Zakharov,
  %``Particle capture cross-sections for a Reissner-Nordstrom black hole,''
  Class.\ Quant.\ Grav.\  {\bf 11}, 1027 (1994).
  %%CITATION = CQGRD,11,1027;%%

%\cite{Zakharov:2005ek}
\bibitem{Zakharov:2005ek}
  A.~F.~Zakharov, F.~De Paolis, G.~Ingrosso and A.~A.~Nucita,
  %``Direct Measurements of Black Hole Charge with Future Astrometrical
  %Missions,''
  New Astron.\  {\bf 10}, 479 (2005)
  [arXiv:astro-ph/0505286].
  %%CITATION = NEWAS,10,479;%%

%\cite{Schee:2008kz}
\bibitem{Schee:2008kz}
  J.~Schee and Z.~Stuchlik,
  %``Optical phenomena in brany Kerr spacetimes,''
  arXiv:0810.4445 [astro-ph].
  %%CITATION = ARXIV:0810.4445;%%

%\cite{Bambi:2008jg}
\bibitem{Bambi:2008jg}
  C.~Bambi and K.~Freese,
  %``Apparent shape of super-spinning black holes,''
  Phys.\ Rev.\  D {\bf 79}, 043002 (2009)
  [arXiv:0812.1328 [astro-ph]].
  %%CITATION = PHRVA,D79,043002;%%

\bibitem{Penrose} R.~Penrose, Ann.\ N.\ Y.\ Acad.\ Sci.\ {\bf 224}, 125 (1973).

%\cite{Virbhadra:2002ju}
\bibitem{Virbhadra:2002ju}
  K.~S.~Virbhadra and G.~F.~R.~Ellis,
  %``Gravitational lensing by naked singularities,''
  Phys.\ Rev.\  D {\bf 65}, 103004 (2002).
  %%CITATION = PHRVA,D65,103004;%%

%\cite{Nakao:2002kc}
\bibitem{Nakao:2002kc}
  K.~Nakao, N.~Kobayashi and H.~Ishihara,
  %``How Does Naked Singularity Look?,''
  Phys.\ Rev.\  D {\bf 67}, 084002 (2003)
  [arXiv:gr-qc/0211061].
  %%CITATION = PHRVA,D67,084002;%%
  
%\cite{Virbhadra:2007kw}
\bibitem{Virbhadra:2007kw}
  K.~S.~Virbhadra and C.~R.~Keeton,
  %``Time delay and magnification centroid due to gravitational lensing by black
  %holes and naked singularities,''
  Phys.\ Rev.\  D {\bf 77}, 124014 (2008)
  [arXiv:0710.2333 [gr-qc]].
  %%CITATION = PHRVA,D77,124014;%%
  
%\cite{Gyulchev:2008ff}
\bibitem{Gyulchev:2008ff}
  G.~N.~Gyulchev and S.~S.~Yazadjiev,
  %``Gravitational Lensing by Rotating Naked Singularities,''
  Phys.\ Rev.\  D {\bf 78}, 083004 (2008)
  [arXiv:0806.3289 [gr-qc]].
  %%CITATION = PHRVA,D78,083004;%%
  
%\cite{Falcke:1999pj}
\bibitem{Falcke:1999pj}
  H.~Falcke, F.~Melia and E.~Agol,
  %``Viewing the Shadow of the Black Hole at the Galactic Center,''
  [Astrophys.\ J.\  {\bf 528}, L13 (2000)]
  arXiv:astro-ph/9912263.
  %%CITATION = ASTRO-PH/9912263;%%
  
%\cite{Takahashi:2004xh}
\bibitem{Takahashi:2004xh}
  R.~Takahashi,
  %``Shapes and Positions of Black Hole Shadows in Accretion Disks and Spin
  %Parameters of Black Holes,''
  J.\ Korean Phys.\ Soc.\  {\bf 45}, S1808 (2004)
  [Astrophys.\ J.\  {\bf 611}, 996 (2004)]
  [arXiv:astro-ph/0405099].
  %%CITATION = ASJOA,611,996;%%

%\cite{Beckwith:2004ae}
\bibitem{Beckwith:2004ae}
  K.~Beckwith and C.~Done,
  %``Extreme Gravitational Lensing near Rotating Black Holes,''
  Mon.\ Not.\ Roy.\ Astron.\ Soc.\  {\bf 359}, 1217 (2005)
  [arXiv:astro-ph/0411339].
  %%CITATION = MNRAA,359,1217;%%
      
%\cite{Broderick:2005at}
\bibitem{Broderick:2005at}
  A.~E.~Broderick and A.~Loeb,
  %``Frequency-Dependent Shift in the Image Centroid of the Black Hole at the
  %Galactic Center as a Test of General Relativity,''
  Astrophys.\ J.\  {\bf 636}, L109 (2006)
  [arXiv:astro-ph/0508386].
  %%CITATION = ASJOA,636,L109;%%
      
%\cite{Broderick:2005xa}
\bibitem{Broderick:2005xa}
  A.~E.~Broderick and R.~Narayan,
  %``On The Nature of the Compact Dark Mass at the Galactic Center,''
  Astrophys.\ J.\  {\bf 638}, L21 (2006)
  [arXiv:astro-ph/0512211].
  %%CITATION = ASJOA,638,L21;%%

%\cite{Takahashi:2007ac}
\bibitem{Takahashi:2007ac}
  R.~Takahashi and K.~Y.~Watarai,
  %``Eclipsing light curves for accretion flows around a rotating black hole
  %and atmospheric effects of the companion star,''
  Mon.\ Not.\ Roy.\ Astron.\ Soc.\  {\bf 374}, 1515 (2007)
  [arXiv:0704.2643 [astro-ph]].
  %%CITATION = MNRAA,374,1515;%%
    
%\cite{Wu:2007bq}
\bibitem{Wu:2007bq}
  S.~M.~Wu and T.~G.~Wang,
  %``Iron line profiles and self-shadowing from relativistic thick accretion
  %discs,''
  Mon.\ Not.\ Roy.\ Astron.\ Soc.\  {\bf 378}, 841 (2007)
  [arXiv:0705.1796 [astro-ph]].
  %%CITATION = MNRAA,378,841;%%
  
%\cite{Huang:2007us}
\bibitem{Huang:2007us}
  L.~Huang, M.~Cai, Z.~Q.~Shen and F.~Yuan,
  %``Black Hole Shadow Image and Visibility Analysis of Sagittarius A*,''
  Mon.\ Not.\ Roy.\ Astron.\ Soc.\  {\bf 379}, 833 (2007)
  [arXiv:astro-ph/0703254].
  %%CITATION = MNRAA,379,833;%%

\bibitem{Young:1976} P.~J.~Young, Phys.\ Rev.\  D {\bf 14}, 3281 (1976).

\bibitem{AdeVries} A. de Vries, Class. Quant. Grav. {\bf 17}, 123 (2000).

%\cite{Hioki:2008zw}
\bibitem{Hioki:2008zw}
  K.~Hioki and U.~Miyamoto,
  %``Hidden symmetries, null geodesics, and photon capture in the Sen black
  %hole,''
  Phys.\ Rev.\  D {\bf 78}, 044007 (2008)
  [arXiv:0805.3146 [gr-qc]].
  %%CITATION = PHRVA,D78,044007;%%

\bibitem{MTW}
C.~W.~Misner, K.~S.~Thorne, and J.~A.~Wheeler, {\it Gravitation}, (Freeman, 1973).

\bibitem{Kerr} R.~P.~Kerr, Phys.\ Rev.\ Lett.\ {\bf 11}, 237 (1963).

\bibitem{Yano} K.~Yano, Ann.\ Math.\ {\bf 55}, 328 (1952).

%\cite{Carter:1968rr}
\bibitem{Carter:1968rr}
  B.~Carter,
  %``Global structure of the Kerr family of gravitational fields,''
  Phys.\ Rev.\  {\bf 174}, 1559 (1968).
  %%CITATION = PHRVA,174,1559;%%

%\cite{Chandrasekhar:1985kt}
\bibitem{Chandrasekhar:1985kt}
  S.~Chandrasekhar,
  {\it The mathematical theory of black holes},
  (Oxford Univ. Press, 1992) p. 646.

%\cite{Bardeen:1972fi}
\bibitem{Bardeen:1972fi}
  J.~M.~Bardeen, W.~H.~Press and S.~A.~Teukolsky,
  %``Rotating Black Holes: Locally Nonrotating Frames, Energy Extraction, And
  %Scalar Synchrotron Radiation,''
  Astrophys.\ J.\  {\bf 178}, 347 (1972).
  %%CITATION = ASJOA,178,347;%%


\end{thebibliography}
\end{document}